\documentclass[aps,pra,twocolumn,reprint]{revtex4-1}
\usepackage[english]{babel}
\usepackage{xcolor,graphicx}
\usepackage{tabularx}

\usepackage{amsmath}
\usepackage{amssymb}
\usepackage{amsbsy}

\usepackage[retainorgcmds]{IEEEtrantools}

\usepackage[normalem]{ulem}

\newcommand{\ud}{\mathrm{d}}
\newcommand{\ui}{\mathrm{i}}

\DeclareMathOperator{\Tr}{Tr}

\DeclareMathOperator{\im}{Im}
\DeclareMathOperator{\re}{Re}
\DeclareMathOperator{\Var}{Var}

\newcommand*\diff{\mathop{}\!\mathrm{d}}
\newcommand*\diffo{\frac{\mathop{}\!\mathrm{d \omega}}{\sqrt{2\pi}}}

\begin{document}


\title{Remote Hamiltonian Interactions Mediated by Light}


\author{Thomas M. Karg}
\email{thomas.karg@unibas.ch}
\affiliation{Department of Physics, University of Basel, Klingelbergstrasse 82, 4056 Basel, Switzerland}

\author{Baptiste Gouraud}
\affiliation{Department of Physics, University of Basel, Klingelbergstrasse 82, 4056 Basel, Switzerland}

\author{Klemens Hammerer}
\affiliation{Institute for Theoretical Physics and Institute for Gravitational Physics (Albert Einstein Institute), Leibniz Universit\"at Hannover, Callinstra\ss e 38, 30167 Hannover, Germany}

\author{Philipp Treutlein}
\email{philipp.treutlein@unibas.ch}
\affiliation{Department of Physics, University of Basel, Klingelbergstrasse 82, 4056 Basel, Switzerland}

\date{\today}

\begin{abstract}
We address a fundamental question of quantum optics: Can a beam of light mediate coherent Hamiltonian interactions between two distant quantum systems? This is an intriguing question whose answer is not \emph{a priori} clear, since the light carries away information about the systems and might be subject to losses, giving rise to intrinsic decoherence channels associated with the coupling. Our answer is affirmative and we derive a particularly simple sufficient condition for the interactions to be Hamiltonian: The light field needs to interact twice with the systems and the second interaction has to be the time reversal of the first. We demonstrate that, even in the presence of significant optical loss, coherent interactions can be realized and generate substantial amounts of entanglement between the systems. Our method is directly applicable for building hybrid quantum systems, with relevant applications in the fields of optomechanics and atomic ensembles.
\end{abstract}


\maketitle

\section{Introduction}
\label{sec:intro}

Light is an excellent carrier of information over a distance. It not only has become an essential tool of modern communication technologies, but is also the most realistic quantum information carrier for large scale quantum communication networks \cite{Gisin2007}. On the other hand, coherent Hamiltonian coupling between quantum objects is typically observed on a local scale and mediated by short-range interactions, e.g. ions interacting via the Coulomb force \cite{Blatt2008} or superconducting qubits via capacitive or inductive coupling \cite{Clarke2008}.

Instead of carrying information from one point to another, light can also be used to mediate a remote Hamiltonian interaction between two distant objects and thus create an ``effective spring'' between them. We present here a formalism to describe such light-mediated interactions, discuss their properties, and in particular derive conditions for them to be Hamiltonian.

Light-mediated interactions not only allow one to remotely couple two similar objects, but any set of different objects, as soon as a proper light-matter interface exists for each of them. This may open up new possibilities for quantum technologies, allowing one to combine the strengths of disparate devices in order to meet the requirements of quantum technologies in a modular setup \cite{Wallquist2009}. 

\begin{figure}[!b]
\centering
\includegraphics[width=8.5cm]{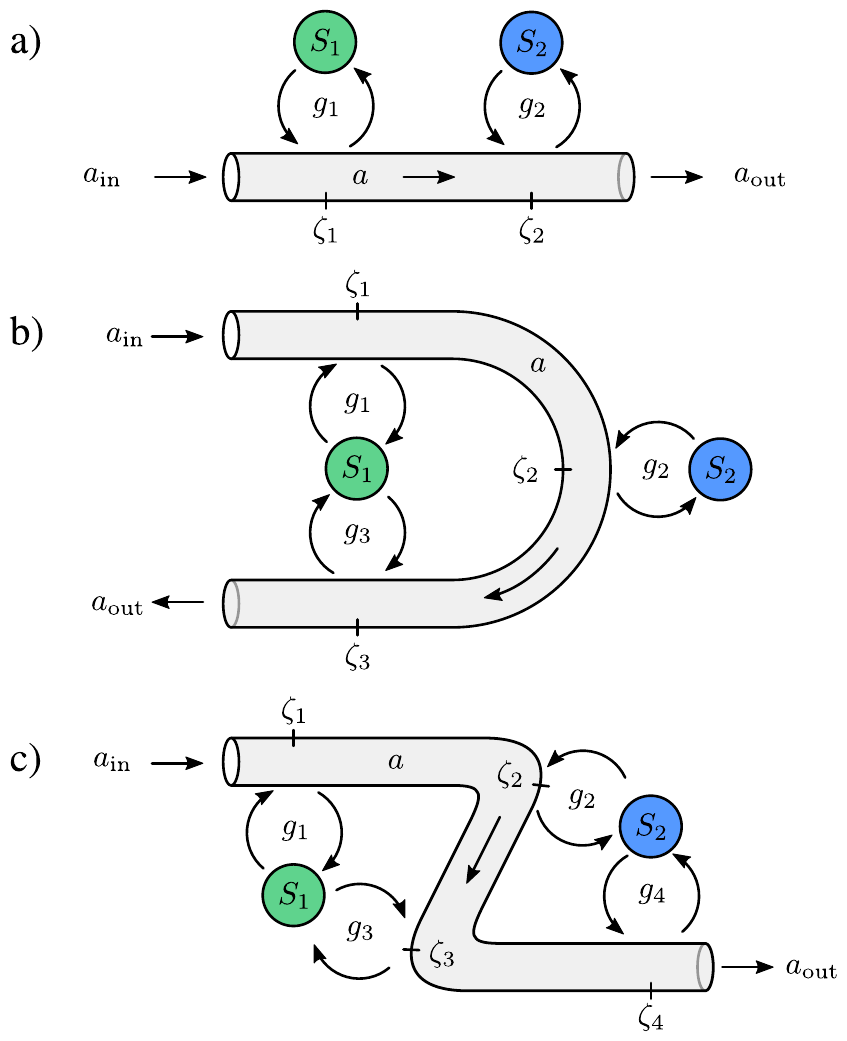}
\caption{Experimental schemes considered in this article. (a) Standard cascaded setup where two systems $S_1$ and $S_2$ interact sequentially with a 1D optical mode $a$ and realize a unidirectional interaction $1\to2$. (b) Looped cascaded setup where system 1 couples to the light field twice, once before system 2 and once after, thus realizing the interaction $1\to2\to1$. (c) Setup with double passes through both systems, realizing the interaction $1\to2\to1\to2$.}
\label{fig:schemes}
\end{figure}

We consider a quite general scenario where quantum systems couple sequentially, and possibly repeatedly, to a one-dimensional (1D) waveguide. Such a setup is described theoretically in the framework of cascaded quantum systems \cite {Gardiner1993, Carmichael1993, Gardiner2004} and generically results in Hamiltonian interactions among the quantum systems along with collective decay at a comparable level. This conceptual framework was applied fruitfully in the description of cascaded optical cavities \cite{Collett1984}, atomic ensembles interacting with light in free space \cite{Muschik2013}, superconducting systems \cite{Lalumiere2013} and optomechanical devices \cite{Stannigel2010}. It also received renewed interest in recent years in the context of chiral quantum optics where near-field effects in nanophotonics are exploited in order to realize unidirectional coupling of quantum emitters to waveguides \cite{Pichler2015}. Our work contributes to the theory of cascaded quantum systems by demonstrating that it is possible to exploit the light-induced interaction for coherent dynamics among the quantum systems by efficiently suppressing the relative strength of light-induced decoherence. The main idea is to use a looped geometry where one or several of the cascaded quantum systems interact with the beam of light twice, effectively reducing or removing decoherence via destructive interference of quantum noise. For the specific case of superconducting systems such an effect has been studied theoretically in Ref.~\cite{Kockum2018}. Here we aim to develop a general framework for the engineering of remote Hamiltonian interactions mediated by light which is applicable to a large variety of cascaded quantum systems.

We focus on simple geometries involving multiple passes of light through two quantum systems $S_1$ and $S_2$ which are sketched in Fig.~\ref{fig:schemes}. In geometry (a), because light carries information in a single direction, the effective dynamics cannot be reduced to a Hamiltonian. In (b) however, where light travels back and forth, the effective interaction can be Hamiltonian and we derive a simple condition for this: The second interaction of light with $S_1$ must be the time reversal of the first. Light necessarily exits the optical mode with some information about the two systems which leads to a diffusive noise process associated with measurement back-action. In configuration (a), the strength of this noise process will always be stronger than the mediated coherent interaction. In case (b), however, engineering a time reversal in the two light-matter interactions with $S_1$ cancels the back-action noise and erases the measurement done by the light field. This allows us to increase the coherent coupling strength without adding excess noise and we show that, in principle, the coherent coupling strength can be made arbitrarily stronger than the light-induced diffusion rate on $S_2$. To go one step further, the remaining back-action noise on $S_2$ can also be removed by extending the simple looped geometry by another time-reversed light-matter interaction with $S_2$ as depicted in Fig.~\ref{fig:schemes}(c). In the absence of any back-action noise, this scheme realizes a perfect Hamiltonian interaction between two quantum systems.

\begin{figure}[!b]
\centering
\includegraphics[width=8.5cm]{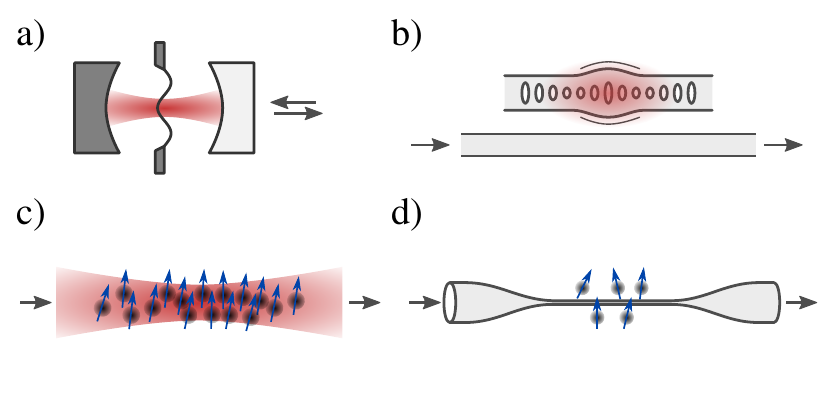}
\caption{Overview of suitable experimental systems to build cascaded systems as shown in Fig.~\ref{fig:schemes}. (a) Membrane-in-the-middle optomechanical cavity coupled to a free-space laser beam \cite{Purdy2013, Nielsen2017}, (b) integrated optomechanical crystal device coupled to an optical waveguide \cite{Safavi-Naeini2014, Riedinger2018}, (c) collective atomic spin ensemble probed by a free-space laser beam \cite{Julsgaard2001}, and (d) atoms coupled to a nano-fiber \cite{Sayrin2015, Solano2017}.}
\label{fig:realizations}
\end{figure}

Previously, the same formalism has been used to treat hybrid mechanical-atomic systems \cite{Hammerer2010a,Vogell2013,Bennett2014,Vogell2015}, lacking, however, precise and general statements about the role of optical losses, optical back-action noise, and the time-reversal condition required to achieve Hamiltonian dynamics. Here, we address all of these open questions in a unified framework, thus greatly facilitating the design of future experiments.

Our scheme readily applies to a variety of quantum systems that interact coherently with free-space or guided light (for examples, see Fig.~\ref{fig:realizations}), in particular optomechanical systems \cite{Aspelmeyer2014}, atomic ensembles \cite{Hammerer2010}, nanophotonic devices \cite{Lodahl2017, Chang2018}, and hybrid quantum systems thereof \cite{Treutlein2014, Kurizki2015}. These systems exhibit large cooperativity for the coupling to the waveguide mode as compared to all other modes.

A number of  related works that are close to but beyond the scope of this article also exist and can be discussed and interpreted with the insight from the present article. This includes single-pass entanglement schemes with conditioning on a measurement of the output field \cite{Krauter2011, Muschik2013, Moller2017, Huang2018}, or experiments involving cavity-mediated effective interactions \cite{Ritsch2013, Spethmann2015}. We remark that the results presented here could be generalized to describe light-mediated dynamics in optical ring cavities. We emphasize, however, that the free-space character of our scheme is particularly appealing for high-bandwidth and long-distance networks, and allows local operations on the optical field between nodes which can be used to modify the character of the interaction on the timescale of the mediated dynamics.

This article is organized as follows. In Sec.~\ref{sec:general}, we consider the general problem of a set of isolated quantum objects interacting locally and possibly repeatedly with a traveling light field. The field carries information between the different objects, creating an effective interaction, before exiting the system. The local light-matter interactions are assumed to be Hamiltonian and linear in the field quadratures. Propagation delays are neglected relative to the local and effective interaction dynamics. For this problem, we derive a general Markovian master equation that captures the effective dynamics.

In Sec.~\ref{sec:specificgeometries}, we apply the results of the general theory to the different geometries of Fig.~\ref{fig:schemes} and discuss the resulting dynamics.
Based on a decomposition of the master equation into Hamiltonian and dissipative evolution, we identify conditions such that the effective dynamics is dominated by the Hamiltonian term. We find that in these cases light-induced dissipation can in principle be made arbitrarily small such that the effective coupling becomes fully coherent. 

Section~\ref{sec:dynamics} discusses the cooperativity as a figure of merit for coherent dynamics and analyzes different applications relevant for hybrid quantum systems. Straightforward results also arise for a scenario with multiple passes of light through the same object. This leads, for example, to an apparent cancellation of radiation-pressure noise in an optomechanical system or deterministic squeezing in a spin ensemble.

\section{General Description}
\label{sec:general}

\begin{figure}[!t]
\centering
\includegraphics[width=8.5cm]{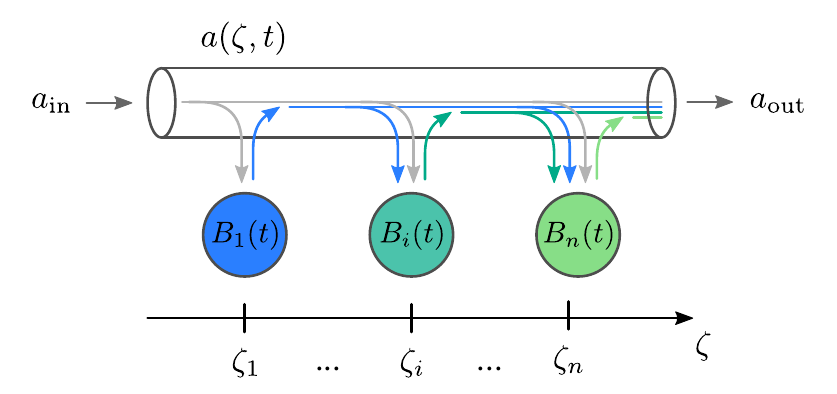}
\caption{Sketch of the cascaded light-matter interactions.}
\label{fig:cascade}
\end{figure}

We consider $N$ quantum systems that sequentially interact with a common traveling electromagnetic field mode $a$. The path of the electromagnetic mode is parametrized by a position coordinate $\zeta$. Each of the interactions happens at a distinct spatial coordinate $\zeta_j$ along the optical path and couples a system operator $B_j$ to the local field $a(\zeta_j)$ with coupling strength $g_j\in\mathbb{R}$ (see Fig.~\ref{fig:cascade}). A total of $n\geq N$ interactions are allowed such that a system can interact with the field more than once. We work in a rotating frame for the optical mode where the full Hamiltonian reads
\begin{eqnarray}
H &=& H_0 + H_\mathrm{int},\\
H_0 &=& \sum_{i=1}^{N} H_i + \int \diff{\omega} \hbar \omega \;a^\dag(\omega)a(\omega),\\
H_\mathrm{int} &=& \sum_{j=1}^{n} \hbar g_j \left(B_j^\dag a(\zeta_j) + a^\dag(\zeta_j) B_j \right).\label{eq:H_int_general}
\label{eq:Hamiltonian1}
\end{eqnarray}
The coordinates $\zeta_j$ are chosen in increasing order such that they can be associated with propagation times $\tau_j = \zeta_j/c$, where $c$ is the speed of light. Delays between interactions $j$ and $k$ are denoted $\tau_{jk} = \tau_j - \tau_k$. 

The operators $B_j$ can be arbitrary operators acting on a single system. However, their typical form for harmonic oscillators or spin systems as considered in this work is $B_j = e^{\ui\phi_j}(\mu_j b_{s_j} + \nu_j b_{s_j}^\dag)$. Here, $b_{s_j}$ and $b_{s_j}^\dag$ are annihilation and creation operators, respectively, of an oscillator or ladder operators of a spin \cite{Pezze2018} satisfying the commutation relation $[b_{j}, b_{k}^\dag] = \delta_{jk}$. We use the label $s_j$ for the system that is involved in light-matter interaction $j$. 
The phase $\phi_j$ selects a specific optical quadrature and the coefficients $\mu_j = \cos(\theta_j)$ and $\nu_j = \sin(\theta_j)$ correspond to different amplitudes for Stokes and anti-Stokes scattering, respectively, realizing beam-splitter and parametric gain interactions with the light field \cite{Kraus2003, Muschik2011}. 
The parameters $\phi_j$ and $\theta_j$ can be tuned experimentally. For light-matter interactions based on two-photon transitions involving a classical drive, $\phi_j$ is the relative phase between the quantum and classical fields. It is adjustable via polarization optics or interferometry.
In cavity-optomechanical systems, tuning the scattering amplitudes $\mu_j, \nu_j$ is commonly achieved via the detuning of the pump laser relative to the cavity resonance \cite{Aspelmeyer2014}. For atomic spin ensembles it requires adjusting the pump laser's polarization and detuning relative to the atomic transition \cite{Hammerer2010}.
Note that we assume $B_j$ to be dimensionless such that $g_j^2$ has dimension Hz and can be interpreted as the measurement rate with which information about $B_j$ is read out by the light field \cite{Clerk2010}. 
The local Hamiltonians considered here are those for harmonic oscillators, i.e. $H_i = \hbar \Omega_i b_i^\dag b_i$ with oscillation frequency $\Omega_i$.

We remark that for linearized light-matter interactions as typically encountered in cavity optomechanics or quantum optics with atomic ensembles the coupling strengths $g_j$ are proportional to the field amplitude of a pump laser co-propagating with the quantum field. In fact, it is the pump laser that enhances the coupling to a single mode of the waveguide over that to all other modes. In chiral quantum optics, such uni-directional light-matter interactions can also be engineered without the need of a pump laser. 


The local annihilation operator for the light field is defined as
\begin{equation}
a(\zeta) = \int \diffo a(\omega) e^{\ui\omega \zeta/c},
\label{eq:field_fourier_transform}
\end{equation}
where it is implicitly assumed that the dynamics are limited to a small bandwidth, i.e. sidebands around the carrier frequency of the laser.

In the following we derive equations for the effective coupled dynamics of the N quantum systems by eliminating the light field in a Born-Markov approximation. In \ref{sec:HLEquations} we write the Heisenberg-Langevin equations of motion in the spirit of the input-output formalism \cite{Gardiner1985} commonly used in quantum optics, cavity optomechanics and cavity quantum electrodynamics. They provide insight on how one system drives another via the light field and can be used to obtain a master equation with stochastic differential calculus \cite{Gardiner2004}. In Sec.~\ref{sec:MasterEquation} we directly derive such a master equation by tracing out the light field within the density matrix formalism. Losses are then included and the resulting coupled dynamics are later discussed for the different geometries of Fig. 1.

\subsection{Heisenberg-Langevin equations}
\label{sec:HLEquations}
In the Heisenberg picture, the equation of motion of the optical field is
\begin{equation}
\dot{a}(\omega,t) = -\ui \omega a(\omega,t) -\ui \sum_{j=1}^{n}\frac{g_j}{\sqrt{2\pi}} B_j(t)  e^{-\ui \omega \tau_j}.
\end{equation}
This equation of motion is subject to the initial condition $a(\omega,t=0) = a_{0}(\omega)$. Formal integration to a time $t$ larger than all propagation delays $\tau_{jk}$ and a Fourier transform yields \cite{Gardiner1985, Gardiner1993}


\begin{equation}
a(\zeta,t) =  a_\mathrm{in}(\zeta,t) -\ui \sum_{j=1}^{n} g_j  B_j(t-(\zeta-\zeta_j)/c)  \Theta(\zeta-\zeta_j).
	\label{eq:field1}
\end{equation}
Here, $a_\mathrm{in}(\zeta)$ is the Fourier transform of $a_0(\omega)$ according to Eq.~\eqref{eq:field_fourier_transform} which is the input field driving the system. In practice, $a(\zeta,t) = a_\mathrm{in}(\zeta,t)$ for $\zeta<\zeta_1$. The Heaviside step function is defined by $\Theta(x) = 1$ for $x>0$, $\Theta(x) = 0$ for $x<0$ and $\Theta(0) = 1/2$.
Evaluating the above expression \eqref{eq:field1} at positions $\zeta > \zeta_n$ yields the output field
\begin{equation}
a_\mathrm{out}(t) = a_\mathrm{in}(t) - \ui \sum_{j=1}^{n} g_j  B_j(t+\tau_j),
\end{equation}
which we have defined as $a_\mathrm{out}(t)=a(\zeta,t+\zeta/c)$ and the input field via $a_\mathrm{in}(t)=a_\mathrm{in}(\zeta,t+\zeta/c)=a_\mathrm{in}(0,t)$.


The time evolution of the operator $b_i$ of system $i$ interacting with the optical mode via Eq.~\eqref{eq:H_int_general} is
\begin{equation}
\dot{b}_i = \mathcal{L}_i b_i -\ui\sum_{j=1}^{n} g_j  \left([b_i,B_j^\dag] a(\zeta_j) + a^\dag(\zeta_j) [b_i,B_j]\right),
\end{equation}
with local dynamics captured by a Liouvillian $\mathcal{L}_i$ that includes dynamics due to $H_0$. Inserting expression \eqref{eq:field1} gives
\begin{eqnarray}
\dot{b}_i &=& \mathcal{L}_i b_i -\ui\sum_{j=1}^{n} g_j  \left([b_i,B_j^\dag] a_\mathrm{in}(\zeta_j) + a_\mathrm{in}^\dag(\zeta_j) [b_i,B_j]\right)\nonumber\\
	&& -  \sum_{j=1}^{n}\sum_{k\leq j} g_j g_k  \Theta(\zeta_j-\zeta_k) \Big([b_i,B_j^\dag(t)] B_k(t-\tau_{jk})  \nonumber\\
	& & \qquad \qquad \qquad \qquad - B_k^\dag(t-\tau_{jk}) [b_i,B_j(t)]\Big).
\end{eqnarray}
This expression is one of our main results. It can be divided into three parts, (i) internal dynamics, (ii) source terms of the input field driving the systems and (iii) interactions between systems. The fact that the optical input field drives all systems in a similar way means that the resulting noise processes are correlated between all systems. It has been demonstrated that these noise channels can be made to destructively interfere in the collective measurement 
of two oscillators with equal and opposite linear responses \cite{Moller2017}. If the quantum noise correlations induced by the input field are stronger than intrinsic system noise processes the collective measurement can establish entanglement or even Einstein-Podolsky-Rosen (EPR) correlations \cite{Hammerer2009}. In this paper we focus on the direct system-system interactions that can be harnessed to generate coherent quantum dynamics and unconditional quantum correlations. A particular aim of this paper is to explore the conditions under which the coherent mediated interaction can compete against the quantum noise added by the light field. To gain further insight into the interactions achievable within this framework we must make assumptions on the topology of the optical path and the form of the local interactions.

\subsection{Master equation}
\label{sec:MasterEquation}

\subsubsection{Derivation}

Following \citet{Gardiner2004} an alternative description of the effective dynamics can be obtained in the framework of a quantum optical master equation. We take the perspective that the optical mode is a vacuum bath to which all systems couple in a time-ordered fashion. To derive the master equation we work in an interaction frame with respect to the Hamiltonian $H_0$. Operators in the interaction frame are marked with a tilde symbol. The time evolution of the reduced density operator $\rho = \Tr_L\{\rho_\mathrm{tot}\}$ of the systems 1 to N is obtained by tracing out the light field $L$. This gives
\begin{equation}
\dot{\tilde{\rho}}(t) = -\frac{1}{\hbar^2} \int_0^{t} \Tr_L\left\{[\tilde{H}_\mathrm{int}(t),[\tilde{H}_\mathrm{int}(t'),\tilde{\rho}_\mathrm{tot}(t')]]\right\} \diff{t'}.
\label{eq:MasterEquation0}
\end{equation}
We then make a weak-coupling and Markov approximation \cite{Gardiner2004}. This replaces the full density matrix $\rho_\mathrm{tot}(t')$ in Eq.~\eqref{eq:MasterEquation0} by $\rho(t)\otimes \rho_{L,0}$ and extends the lower limit of the integral to $-\infty$. The state $\rho_{L,0}$ of the optical mode is the vacuum state such that the only non-vanishing optical correlation function is $\Tr_L\left\{ a(\omega) a^\dag(\omega')\rho_{L,0}\right\} = \delta(\omega-\omega')$. Physically, we assume that light exits the cascaded systems on a timescale that is fast when compared to the system dynamics and is only weakly perturbed by the light-matter interaction. By virtue of these approximations we can derive a master equation of the form
\begin{equation}
\dot{\tilde{\rho}} = -A\tilde{\rho} - \tilde{\rho} A^\dag + \mathcal{J}\tilde{\rho},
\label{eq:master_eq_general_form}
\end{equation}
where
\begin{eqnarray}
A	&=& \sum_{j}\sum_{k<j} g_j g_k  \tilde{B}_j^\dag(t) \tilde{B}_k(t - \tau_{jk}) \label{eq:a_operator_1}\\
	& & \quad +\; \sum_{j} \frac{g_j^2}{2}  \tilde{B}_j^\dag(t) \tilde{B}_j(t),
	\label{eq:a_operator_2}
\end{eqnarray}
and
\begin{eqnarray}
\mathcal{J}\tilde{\rho}	&=& \sum_{j}\sum_{k<j} g_j g_k \tilde{B}_k(t - \tau_{jk})\tilde{\rho}(t) \tilde{B}_j^\dag(t) + \;\mathrm{h.c.} \label{eq:j_operator_1}\\
& & + \sum_{j} g_j^2 \tilde{B}_j(t)\tilde{\rho}(t) \tilde{B}_j^\dag(t).
\label{eq:j_operator_2}
\end{eqnarray}
We remark that this result also holds in the case where the coupling constants $g_i$ or the phase factors $\phi_i$ or $\theta_i$ determining the local interactions are time dependent. In this case these parameters are evaluated at the same times as their parent system operators $B_i$. 

The structure of the general master equation derived above demands some explanation. Looking at the expression for the operators $A$ and $\mathcal{J}$, we distinguish between two types of contributions: (i) lines \eqref{eq:a_operator_1} and \eqref{eq:j_operator_1} describe correlated dynamics mediated by the light field. Any system $s_j$ is driven by other systems $s_k$ with $k<j$ that were probed by the light field at earlier times. Causality is preserved because interactions with systems probed in the future ($k>j$) are not present. The coupling constants for these interactions are the products $g_j g_k$ of the coupling strengths of the individual light-matter interactions. We note that these correlated dynamics can be of either dissipative or unitary character, i.e., collective damping and amplification or Hamiltonian interaction. (ii) Lines \eqref{eq:a_operator_2} and \eqref{eq:j_operator_2} contain purely non-unitary time evolution acting on the individual systems with corresponding dissipation rates $g_j^2$. This results in radiative decay as in spontaneous emission or decay of an optical cavity \cite{Gardiner1985} and associated diffusion due to quantum noise from the input field. Since these noise processes are uncorrelated, they destroy quantum coherence between the systems.

In order to harness the mediated interactions for inter-system entanglement and coherent dynamics, they have to be made stronger than the uncorrelated quantum noise. At first sight this task appears impossible because the coherent coupling strengths $g_j g_k$ can never exceed both dissipation rates $g_j^2$ and $g_k^2$.
However, as we will show in the following section, one can engineer the system-reservoir interaction in order to suppress quantum noise while preserving the effective light-mediated interaction.

\subsubsection{Effective interaction}
To interpret the general master equation \eqref{eq:master_eq_general_form}, we compare it with the Lindblad form
\begin{equation}
\dot{\rho} = -\frac{\ui}{\hbar}[H_\mathrm{eff},\rho] + \sum_k \mathcal{D}[j_k]\rho,
\label{eq:master_eq_general_lindblad}
\end{equation}
with effective Hamiltonian $H_\mathrm{eff}$ and jump operators $j_k$. The Lindblad terms read $\mathcal{D}[j]\rho = j \rho j^\dag - \frac{1}{2}\{ j^\dag j, \rho\}$. Here and in what follows we neglect the time delays $\tau_j$ in accordance with the Markov approximation. We also transform back to the laboratory frame and drop the tilde on top of interaction frame operators. The effective Hamiltonian is then
\begin{eqnarray}
H_\mathrm{eff} &=& \frac{\hbar}{2\ui}(A-A^\dag),
\label{eq:general_eff_hamiltonian}
\end{eqnarray}
and the dissipative part can be written as
\begin{equation}
\sum_k j_k^\dag j_k = A + A^\dag =: \Lambda_\mathrm{eff}.
\label{eq:def_lambdaeff}
\end{equation}
As shown in Appendix~\ref{sec:appendix_master_equation}, the form of $\mathcal{J}\rho = \sum_k j_k \rho j_k^\dag$ is closely linked to that of $\Lambda_\mathrm{eff}$ and it is sufficient to know $A$ or $\Lambda_\mathrm{eff}$ in order to write down the equations of motion. In the model presented so far, the effective Hamiltonian is
\begin{eqnarray}
H_\mathrm{eff} &=& \sum_{j}\sum_{k<j} \hbar g_j g_k  \frac{1}{2\ui}\left(B_j^\dag B_k - B_k^\dag B_j\right),
\end{eqnarray}
and the dissipative dynamics are governed by a single collective jump process $\Lambda_\mathrm{eff} = j_+^\dag j_+$ with jump operator
\begin{eqnarray}
j_+ &=& \sum_j g_j B_j,
\end{eqnarray}
which is a superposition of all subsystem operators. More diverse dissipative dynamics are observed when optical losses are included.

\subsubsection{Master equation including losses} 
It is essential to take into account optical losses in our model, as they will contribute significantly to decoherence by introducing uncorrelated vacuum noise. To describe losses we insert beam splitters with (amplitude) transmission coefficient $\eta_j$ between every pair of interactions $j$ and $j+1$. The beam-splitter relations
\begin{equation}
a(\zeta_j) \;\to\; \eta_j a(\zeta_j) + \sqrt{1-\eta_j^2} \; h_j(\zeta_j),
\end{equation}
mix the optical mode with an uncorrelated mode $h_j$ in the vacuum state. With losses the new time evolution operator becomes
\begin{equation}
A = \sum_{j}\sum_{k<j} \eta_{jk} g_j g_k  B_j^\dag B_k + \sum_{j} \frac{g_j^2}{2}  B_j^\dag B_j,
\label{eq:a_operator_loss}
\end{equation}
where $\eta_{jk} = \eta_k \cdot \ldots \cdot \eta_{j-1}$ is the transmittance from system $k$ to system $j$. The sandwich term changes accordingly:
\begin{equation}
\mathcal{J}\rho = \sum_{j}\sum_{k<j} \eta_{jk} g_j g_k ( B_k \rho B_j^\dag + B_j \rho B_k^\dag) + \sum_{j} g_j^2  B_j \rho B_j^\dag.
\label{eq:sandwich_loss}
\end{equation}

If the coupling constants $g_j$ depend on the amplitude of a co-propagating pump field, they also need to be rescaled with the total transmission until system $j$, i.e. $\eta_{j1} = \eta_1 \cdot \ldots \cdot \eta_{j-1}$. This renormalizes the coupling constants and only becomes important in the case when a system interacts multiple times with the optical mode. 

From the two equations \eqref{eq:a_operator_loss} and \eqref{eq:sandwich_loss}, we see that losses between two systems only affect the cross-coupling terms, but leave the noise terms unchanged. Put another way, the effective interaction mediated by light is weakened relative to the quantum noise added by the light. The Lindblad jump operators in this new setting can be derived by diagonalizing the Hermitian matrix $\Lambda_\mathrm{eff}$ in the basis of the $B_j$ operators. The eigenvalues of $\Lambda_\mathrm{eff}$ are the corresponding damping rates. In the presence of losses there is more than one jump operator with non-zero eigenvalue. In Appendix~\ref{sec:proof_lambdaeff_positive}, we provide a proof that $\Lambda_\mathrm{eff}$ is always positive semidefinite for the master equation derived above, which ensures that it can be written in Lindblad form with positive rates and that the dynamics are completely positive \cite{Vega2017}.

\section{Specific geometries}
\label{sec:specificgeometries}

Having established a general theoretical framework for cascaded quantum systems with looped interactions we now analyze this model for the specific geometries displayed in Fig.~\ref{fig:schemes}.

\subsection{Two objects: Single pass}

\begin{figure}[!b]
\centering
\includegraphics[width=8.5cm]{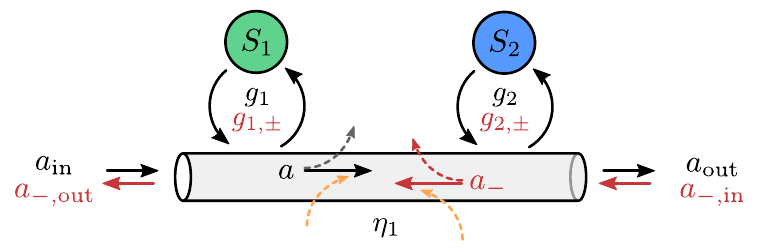}
\caption{Detailed schematic of the single-pass and double-pass coupling schemes. The counter-propagating mode $a_{-}$ is relevant only for the double-pass scheme.}
\label{fig:scheme_12}
\end{figure}

In the case of two cascaded systems like in Figs.~\ref{fig:schemes}(a) and\ref{fig:scheme_12}, 
the effective Hamiltonian is
\begin{equation}
H_\mathrm{eff} = \hbar \eta_1 g_1 g_2 \frac{1}{2\ui}\left(B_2^\dag B_1  - B_1^\dag B_2\right),
\label{eq:h_eff_singlepass}
\end{equation}
and the effective dissipation reads
\begin{equation}
\Lambda_\mathrm{eff} = g_1^2 \; B_1^\dag B_1 + g_2^2  B_2^\dag B_2  + \eta_1 g_1 g_2 \left(B_2^\dag B_1 + B_1^\dag B_2 \right).
\label{eq:lambda_eff_singlepass}
\end{equation}
We note that the interaction terms in Hamiltonian and collective dissipation are out of phase. In the master equation, both terms partially cancel such that only an interaction term proportional to $B_2^\dag B_1$ remains. This is a causality statement which reflects the unidirectional nature of the setup. It means that only system 1 can drive system 2, but not vice versa. The jump operators for this cascaded system are $j_\pm = \sqrt{1\pm \eta_1} (g_1 B_1 \pm g_2 B_2)$ representing dark ($j_-$) and bright modes ($j_+$) of the cascaded system. The effective Hamiltonian mixes these modes  as $H_\mathrm{eff} \propto \ui(j_+^\dag j_- - j_-^\dag j_+)$.

There is an extensive amount of work on exploiting the mediated interaction between two cascaded quantum systems for a state transfer from system 1 to system 2 \cite{Cirac1997, Stannigel2010}. These proposals make use of the effective interaction to transfer an excitation from system 1 to system 2 via a dark state of the cascaded system. By ensuring that the system always stays in the dark mode $j_-$, for which the collective decay rate is suppressed by a factor $1-\eta_1$, unity transfer efficiency can be achieved in principle.

\subsection{Two objects: Double pass}

In order to make the interaction bidirectional, one could exploit a counter-propagating optical mode as sketched in Fig.~\ref{fig:scheme_12} to achieve coupling from system 2 to system 1. For simplicity, we neglect standing wave effects here and assume the counter-propagating mode $a_{-}$ to be independent of the forward-propagating mode $a_+ = a$. Since the two modes are uncorrelated one can simply add up the two resulting effective Hamiltonian and dissipative terms. Because of the antisymmetry of Hamiltonian~\eqref{eq:h_eff_singlepass} under permutation of the systems 1 and 2 we get
\begin{equation}
H_\mathrm{eff} = \hbar \eta_1 (g_{+} - g_{-}) \frac{1}{2\ui}\left(B_2^\dag B_1  - B_1^\dag B_2\right),
\end{equation}
where $g_\pm = g_{1,\pm} g_{2,\pm}$ are the coupling strengths of the light-mediated coupling in forward ($+$) and backward ($-$) directions with coupling strengths of the individual systems to the two modes denoted by $g_{i,\pm}$. The effective dissipation \eqref{eq:lambda_eff_singlepass} is symmetric under permutation of systems 1 and 2 such that with two passes
\begin{eqnarray*}
\Lambda_\mathrm{eff} &=& \sum_i (g_{i,+}^2 + g_{i,-}^2) B_i^\dag B_i \\
& & + \; \eta_1 (g_{+} + g_{-}) \left(B_2^\dag B_1 + B_1^\dag B_2 \right).
\end{eqnarray*}
Consequently, if one naively sets the backward interaction to be of equal strength and phase as the forward interaction, one is left with $H_\mathrm{eff} = 0$ and $\Lambda_\mathrm{eff}$ being twice that of the single-pass scheme, rendering the interaction completely dissipative. In order to still get non-vanishing coupling, one has to implement a coupling that inverts the sign of the backward interaction relative to the forward interaction, e.g., by setting $g_{1, -} = - g_{1, +}$ but $g_{2, -} = g_{2, +}$. This means that the backward interaction is the time reversal of the forward interaction. We remark that this can be achieved naturally if system 1 couples to the photon momentum which is inverted under reflection. In general, as outlined in the beginning of Sec.~\ref{sec:general}, this sign reversal requires appropriate phase shifts to be applied to the optical field between the two systems. 


However, because there are now two independent optical noise inputs the single system decay terms $\sim 2 g_1^2 B_1^\dag B_1 + 2 g_2^2 B_2^\dag B_2$ still remain at twice the original strength. Consequently, the coherent coupling with strength $g = 2 g_1 g_2$ will never exceed both back-action rates $\Gamma_1 = 2 g_1^2$ and $\Gamma_2 = 2 g_2^2$ as outlined before. The only way to suppress quantum noise from the inputs is by recycling the output of the forward propagating optical field as the input for the backward propagating field by placing a mirror after system 2. In that way, noise from $a_+$ is correlated with noise from $a_-$ such that their effect on system 1 cancels because of the equal and opposite coupling strengths. This means that the remaining dissipation $\Lambda_\mathrm{eff} = 2 g_2^2 B_2^\dag B_2$ affects system 2 alone. The quantum noise or back-action cancellation on system 1 now enables us to increase the effective coherent coupling strength above the induced decay rate on system 2 by making $g_1$ much larger than $g_2$. Such a setup has been proposed \cite{Hammerer2010a, Vogell2013, Bennett2014} and experimentally realized \cite{Camerer2011, Joeckel2015, Christoph2018} for atoms coupled to an oscillating mirror. In previous proposals, the importance of back-action cancellation on the atomic ensemble has not been recognized entirely.

\subsection{Two objects: Loop on system 1}
\label{sec:looped_interaciton}
\begin{figure}[!t]
\centering
\includegraphics[width=8.5cm]{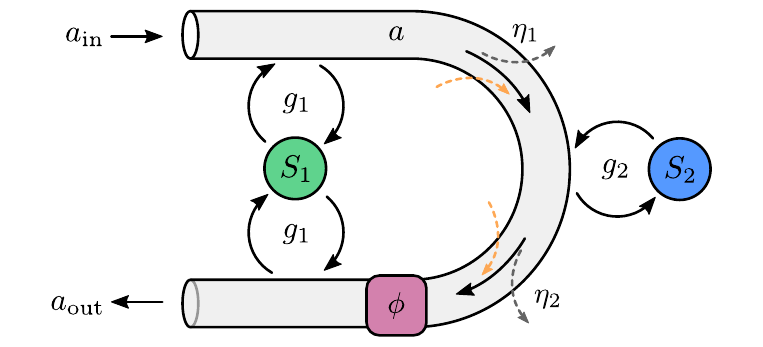}
\caption{Detailed schematic of the coupling scheme involving a loop on system 1.}
\label{fig:scheme_121}
\end{figure}

In order to generalize the double-pass interaction from the previous section we assume two objects coupling to the optical mode in a looped configuration as shown in Figs.~\ref{fig:schemes}(b) and \ref{fig:scheme_121}. Starting from the general expression \eqref{eq:a_operator_loss}, we set $B_3 = B_1 e^{\ui\phi}$ and $g_3=g_1$. The phase shift $\phi$ is motivated by the discussion of constructive and destructive interference of Hamiltonian interaction in the preceding paragraph. It can readily be implemented by local unitary operations on the optical field between interactions with the systems. Applying this to the general expression gives
\begin{eqnarray}
A &=& g_1^2 (1 + \eta_1 \eta_2 e^{-\ui\phi})\; B_1^\dag B_1 + \frac{g_2^2}{2}  B_2^\dag B_2 \\
& & +\; g_1 g_2 (\eta_1  B_2^\dag B_1 + \eta_2 e^{-\ui\phi}B_1^\dag B_2).
\label{eq:a_operator_loop}
\end{eqnarray}


We write the full master equation as
\begin{eqnarray}
\dot{\rho} &=& -\frac{\ui}{\hbar}[H_\mathrm{eff},\rho] + \mathcal{L}\rho,
\label{eq:master_eq_loop}\\
\mathcal{L}\rho &=& \Gamma_1\mathcal{D}[B_1]\rho + \Gamma_2\mathcal{D}[B_2]\rho + \mathcal{G}\rho,\label{eq:master_eq_loop_dissipation}
\end{eqnarray}
with effective Hamiltonian
\begin{eqnarray}
H_\mathrm{eff} &=& -\ui \hbar g_1 g_2 \frac{\eta_1 - \eta_2 e^{\ui\phi}}{2} \; B_2^\dag B_1 + \mathrm{h.c.} \nonumber \\
& & -\;  \hbar g_1^2 \eta_1 \eta_2 B_1^\dag B_1 \sin(\phi).
\label{eq:h_eff_121}
\end{eqnarray}
consisting of interaction between $S_1$ and $S_2$ in the first line and a self-interaction of $S_1$ in the second line. The back-action rates for systems 1 and 2 are given by
\begin{eqnarray}
\Gamma_1 &=& 2 g_1^2 (1 + \eta_1 \eta_2 \cos\phi),\\
\Gamma_2 &=& g_2^2.
\end{eqnarray}
Further, the term
\begin{eqnarray*}
\mathcal{G}\rho &=& - \; \frac{1}{2}g_1 g_2 (\eta_1 + \eta_2 e^{\ui\phi}) [B_2^\dag, B_1 \rho] + \mathrm{h.c.}\\
& & - \; \frac{1}{2}g_1 g_2 (\eta_1 + \eta_2 e^{-\ui\phi}) [B_1^\dag, B_2 \rho] + \mathrm{h.c.}
\end{eqnarray*}
describes collective non-Hamiltonian evolution \cite{Hammerer2010}. We remark that Eq.~\eqref{eq:master_eq_loop_dissipation} is not manifestly in Lindblad form, but it can be brought into this form by diagonalization of $\Lambda_\mathrm{eff}$ as outlined in Appendix~\ref{sec:appendix_master_equation}.


In the following, we define the mean transmission \mbox{$\bar{\eta} = (\eta_1 + \eta_2)/2$} and the  transmission imbalance \mbox{$\Delta\eta = \eta_1 - \eta_2$}. Two interesting cases emerge for different choices of the loop phase $\phi$ which are analyzed in the following. If $\phi = \pi$, the two light-matter interactions of the first system are out of phase, which corresponds to a time reversal. The case $\phi = 0$ corresponds to concatenating two cascaded interactions with opposite order. We will show that the former leads to Hamiltonian dynamics while the latter reproduces a simple cascaded system.

\paragraph*{Coherent dynamics.} In the case of $\phi=\pi$, the Hamiltonian reduces to
\begin{equation}
H_\mathrm{eff} =  \frac{\hbar g}{2\ui}\left(B_2^\dag B_1  - B_1^\dag B_2\right),
\end{equation}
which is solely constituted of an interaction between systems 1 and 2 at rate $g = 2\bar{\eta} g_1 g_2$. The self-interaction of system 1 in the second line of Eq.~\eqref{eq:h_eff_121} cancels. The dissipative part of the evolution reads
\begin{eqnarray*}
\Lambda_\mathrm{eff} &=& 2 g_1^2 (1 - \eta_1 \eta_2)\; B_1^\dag B_1 + g_2^2  B_2^\dag B_2  \\
& & + \;\Delta\eta g_1 g_2 \left(B_2^\dag B_1 + B_1^\dag B_2 \right).
\end{eqnarray*}
Here, the measurement back-action noise on system 1, $\Gamma_1 = 2 g_1^2 (1 - \eta_1 \eta_2)$, is partially canceled down to the level of losses between the two interactions. This is directly reflected in the equation of motion
\begin{eqnarray*}
\dot{b}_1 &=& g_1 \mu_1 \left(a_\mathrm{in}(\zeta_1) - \eta_1 \eta_2 a_\mathrm{in}(\zeta_3) + \sqrt{1-\eta_1^2\eta_2^2} h_\mathrm{in}(\zeta_3)\right) \\
& & + g_1 \nu_1 \left(a^\dag_\mathrm{in}(\zeta_1) - \eta_1 \eta_2 a^\dag_\mathrm{in}(\zeta_3) + \sqrt{1-\eta_1^2\eta_2^2} h^\dag_\mathrm{in}(\zeta_3)\right) \\
& & + \ldots
\end{eqnarray*}
where the ellipsis includes coupling to system 2 and internal dynamics. Here, the destructive interference between the primary input field $a_\mathrm{in}$ at the two positions $\zeta_1$ and $\zeta_3$ becomes evident. Losses introduce an additional noise input $h_\mathrm{in}$  which is uncorrelated with $a_\mathrm{in}$. The rates of these two noise inputs add up to the same value $\Gamma_1$ as obtained from the master equation. We note that time delays add a frequency dependent phase shift between $a_\mathrm{in}(\zeta_1)$ and $a_\mathrm{in}(\zeta_3)$ that renders the cancellation imperfect. These effects are missing in the master equation because time delays have been neglected. Within the rotating-wave approximation the effect of time delays can be captured by $a_\mathrm{in}(\zeta_1) - \eta_1 \eta_2 a_\mathrm{in}(\zeta_3) \approx a_\mathrm{in} (1-\eta_1 \eta_2 e^{-\ui\Omega_1\tau_{13}})$ (see Appendix~\ref{sec:time_delays}). Consequently, perfect back-action cancellation requires $\Omega_1 \tau_{13} \ll 1$, as expected.

Destructive interference of the input noise on system 1 goes along with destructive interference of the signal in the output field
\begin{eqnarray*}
a_\mathrm{out} &=& -\ui g_1 (\eta_1 \eta_2 B_1(t-\tau_{13}) - B_1(t)) -\ui \eta_2 g_2 B_2(t-\tau_{23})\\
& & + \eta_1 \eta_2 a_\mathrm{in} + \sqrt{1-\eta_1^2\eta_2^2} h_\mathrm{in}.
\end{eqnarray*}
We see that in the case of $\phi = \pi$, information written onto the light field by system 1 in the first pass is partially erased in the second pass. 

A transmission imbalance in the two light-mediated interactions adds collective dissipation to the dynamics at a rate $\Gamma_{12} = |\Delta\eta| g_1 g_2$ \cite{Hammerer2010}, which is negligible for a symmetric bi-directional coupling scheme with $\eta_1\approx\eta_2$. In this case, the collective dynamics are entirely Hamiltonian and noise is only introduced at the level of the individual systems.

\paragraph*{Dissipative dynamics.}
In the case where $\phi = 0$, the Hamiltonian evolution is strongly suppressed and can be made to vanish exactly if $\eta_1 = \eta_2$. Here,
\begin{equation}
H_\mathrm{eff} = \hbar \Delta\eta g_1 g_2 \frac{1}{2\ui}\left(B_2^\dag B_1  - B_1^\dag B_2\right) 
\end{equation}
and
\begin{eqnarray}
\Lambda_\mathrm{eff} &=& 2 g_1^2 (1 + \eta_1 \eta_2)\; B_1^\dag B_1 + g_2^2  B_2^\dag B_2  \nonumber \\ & &+ 2\bar{\eta} g_1 g_2 \left(B_2^\dag B_1 + B_1^\dag B_2 \right)
\end{eqnarray}

The main difference of the looped configuration as compared to the simple single pass cascaded interaction lies in the purely dissipative nature of the interaction. Even if the operator $B_2^\dag B_1 - B_1^\dag B_2$ is nonzero, one can eliminate the Hamiltonian interaction completely for balanced transmissions $\eta_1 = \eta_2$.

We recover that for full transmission $\bar{\eta}=1$ the effective dynamics are described by a single dissipative process
\begin{equation}
\dot{\rho} = \mathcal{D}[2 g_1 B_1 + g_2 B_2]\rho.
\end{equation}
For the purpose of generating a two-mode squeezed state of two harmonic oscillators via dissipation \cite{Muschik2011a, Vasilyev2013} one chooses the mode $b_2$ to have positive frequency $\Omega_2 = \Omega > 0$, and the mode $b_1$ to have negative frequency $\Omega_1 = -\Omega $. In the interaction picture with regard to $H_0$, we can then write $2 g_1 \tilde{B}_1 + g_2 \tilde{B}_2 = j_{+} e^{-\ui\Omega t} + j_{-} e^{\ui\Omega t}$ where we defined two jump operators $j_{+} = 2 g_1 \mu_1 b_1 + g_2 \nu_2 b_2^\dag$ and $j_{-} = g_2 \mu_2 b_2 + 2 g_1\nu_1 b_1^\dag$. Care has to be taken that both oscillators couple to the same optical field quadrature such that they both experience the same optical input quantum noise, i.e., $\theta_1 = \theta_2$ and $\phi_1 = \phi_2$. This enables collective decay into an entangled state. Making the rotating-wave approximation one obtains the master equation
\begin{equation}
\dot{\rho} \approx \mathcal{D}[j_{+}]\rho + \mathcal{D}[j_{-}]\rho,
\end{equation}
which is equivalent to the master equation in the simple cascaded system. There does not seem to be a clear advantage of the loop geometry in the case of dissipative interaction.

\subsection{Two objects: Loops on both systems}

\begin{figure}[!b]
\centering
\includegraphics[width=8.5cm]{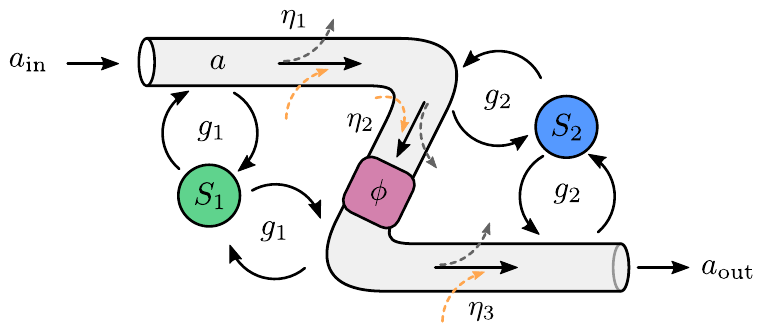}
\caption{Detailed schematic of the double-loop coupling scheme.}
\label{fig:scheme_1212}
\end{figure}

In the looped interaction discussed in the previous section the measurement back-action on system 2 is still present and poses a fundamental limit to the coherence of the remote interaction. We note, however, that this can in principle be remedied by adding another light-matter interaction with system 2 and opposite phase as depicted in Figs.~\ref{fig:schemes}(c) and \ref{fig:scheme_1212} with $B_4 = -B_2$ ($\phi = \pi$). In this second pass through system 2, all information about it will be erased from the light but no further enhancement of the coherent coupling between systems 1 and 2 could be obtained because at this point the light does not contain information about system 1 anymore. This represents the ideal scenario because all back-action is canceled such that $\Lambda_\mathrm{eff} = 0$ and the coherent dynamics are only disturbed by intrinsic damping of the two systems. In a real experiment, there will always be losses in which this scheme possesses an inherent asymmetry because there are three interaction pathways from $S_1$ to $S_2$, but only one from $S_2$ to $S_1$. In the following sections, we will analyze the dynamics that can be realized with both the single loop of Fig.~\ref{fig:schemes}(b) and the double loop.

\section{Coherent dynamics in the looped geometries}
\label{sec:dynamics}

We have seen above that in the looped geometries of Figs.~\ref{fig:schemes}(b) and (c) it is possible to create coupled dynamics which are entirely Hamiltonian. However, decoherence is inherently present as a result of optical back-action noise and also other system-specific decoherence channels will always be present in experiments. This section is devoted to calculating the achievable cooperativity as a figure of merit for coherent dynamics. Beyond that, we also analyze three experimentally relevant applications of our theory.

\subsection{Cooperativity}

In optomechanics and quantum optics, the relevant light-matter interaction strength for system $i$ is the single-pass measurement rate $g_i^2$. Optical cooperativity is commonly defined as the ratio of measurement rate over the intrinsic thermal decoherence rate, $c_i = g_i^2/\gamma_{i,\mathrm{th}}$ for each system $i$ \cite{Hammerer2010, Bowen2015}, referred to as the single-pass cooperativity in the following. The thermal decoherence rate can be expressed as $\gamma_{i,\mathrm{th}} = \gamma_{i}(\bar{n}_i + 1/2)$ with intrinsic damping rate $\gamma_{i}$ and thermal bath occupation $\bar{n}_i$. The contribution $\gamma_i/2$ represents spontaneous scattering.

In a rotating-wave approximation, the dissipative part~\eqref{eq:master_eq_loop_dissipation} of the master equation excluding collective dissipation $\mathcal{G}$ and adding intrinsic decoherence reads
\begin{eqnarray}
\mathcal{L}\rho &=& \sum_i\left[\gamma_{i} (\bar{n}_i + 1) + \mu_i^2 \Gamma_i\right]\mathcal{D}[b_i]\rho\nonumber\\
	& & + \sum_i(\gamma_{i} \bar{n}_i + \nu_i^2 \Gamma_i)\mathcal{D}[b_i^\dag]\rho.\label{eq:lindblad_backaction_rwa}
\end{eqnarray}
This motivates defining a total decoherence rate $\gamma_{i,\mathrm{tot}} = \gamma_{i,\mathrm{th}} + \Gamma_i/2$ covering both intrinsic and light-induced noise processes. The effective Hamiltonian
\begin{equation}
H_\mathrm{eff} = H_\mathrm{BS} +  H_\mathrm{TMS}
\end{equation}
is composed of a beam-splitter (BS) Hamiltonian $H_\mathrm{BS} = \ui \hbar g \alpha (b_1^\dag b_2 - b_2^\dag b_1)$ and a two-mode-squeezing (TMS) Hamiltonian $H_\mathrm{TMS} = \ui\hbar g \beta (b_1 b_2 - b_1^\dag b_2^\dag)$. Here, we have set $\phi_1 = \phi_2 = 0$ for simplicity. The weights are then $\alpha = (\mu_1 \mu_2 - \nu_1 \nu_2)/2 = \cos(\theta_1 + \theta_2)/2$ and $\beta = (\mu_2 \nu_1 - \mu_1 \nu_2)/2 = \sin(\theta_1 - \theta_2)/2$. Coupling is maximized if both oscillators couple to orthogonal optical quadratures, e.g., $\theta_1 = \pi/4 = - \theta_2$ such that $\alpha = \beta = 1/2$. Which one of these two interactions is resonant depends on the oscillator frequencies $\Omega_1$ and $\Omega_2$. While the BS Hamiltonian enables state swaps or the generation of superposition states for $\Omega_1 = \Omega_2$, the TMS Hamiltonian generates non-classical correlations for $\Omega_1 = -\Omega_2$. The following discussion of cooperativity applies to both of them.

The cooperativity $\mathcal{C}$ of the cascaded system compares the strength $g$ of the coherent light-mediated coupling with the intrinsic and light-induced decay rates $\gamma_{i,\mathrm{tot}}$, i.e.,
\begin{equation}
\mathcal{C} = \frac{g^2}{\gamma_{1,\mathrm{tot}} \; \gamma_{2,\mathrm{tot}}}.
\label{eq:cooperativity_0}
\end{equation}
In the following we set $\eta_1 = \eta_2 = \eta$ such that in the looped geometry 1-2-1 we have a coupling constant $g = 2 \eta g_1 g_2$ and the back-action rates are $\Gamma_1 = 2 (1 - \eta^2)g_1^2 $ and $\Gamma_2 = g_2^2$. For zero losses we obtain the asymptotic expression $\mathcal{C} \sim 4 c_1 /(1/2 + 1/c_2)$ which is in principle limited by the single-pass cooperativity of system 1. For finite losses and assuming all individual dissipation rates are small, i.e., large optical cooperativity $c_i \gg 1$, we approximately have
\begin{equation}
\mathcal{C} = \frac{8\eta^2}{1-\eta^2}.
\label{eq:cooperativity_1}
\end{equation}
For linearized couplings as commonly encountered in cavity optomechanics \cite{Aspelmeyer2014} and atom-light interfaces \cite{Hammerer2010} with a single local oscillator experiencing the same losses as the quantum field (i.e. $g_3 = \eta^2 g_1$ and $g_2$ replaced by $\eta g_2$), the rates become $g = (\eta^2+\eta^4) g_1 g_2$, $\Gamma_1 = (1 - \eta^4) g_1^2$, and $\Gamma_2 = \eta^2 g_2^2$. Consequently, we have the scaling
\begin{equation}
\mathcal{C} = \frac{4(\eta^2+\eta^4)}{1-\eta^2}. 
\end{equation}
Note that in this case the interaction is never fully balanced and additional collective dissipation arises at a rate $\Gamma_{12} = 2(\eta^2 - \eta^4) g_1 g_2 $. However, the ratio $g/\Gamma_{12} = (1+\eta^2)/(1-\eta^2) = \mathcal{C}/\eta^2$ is always larger than $\mathcal{C}$. We remark that in principle the loss of the coherent field can be compensated, provided the quantum fields in the sideband frequencies can be separated from the carrier \cite{Moller2017}.

In the double-loop geometry 1-2-1-2 with $g = \eta(3 - \eta^2) g_1 g_2$ and $\Gamma_i = 2 g_i^2 (1 - \eta^2)$, we obtain the loss-limited cooperativity
\begin{equation}
\mathcal{C} = \frac{\eta^2(3 - \eta^2)^2}{(1-\eta^2)^2},
\label{eq:cooperativity_2}
\end{equation}
while for zero losses it is $\mathcal{C} = 4 c_1 c_2$.

\begin{figure}[!t]
\centering
\includegraphics[width=8.5cm]{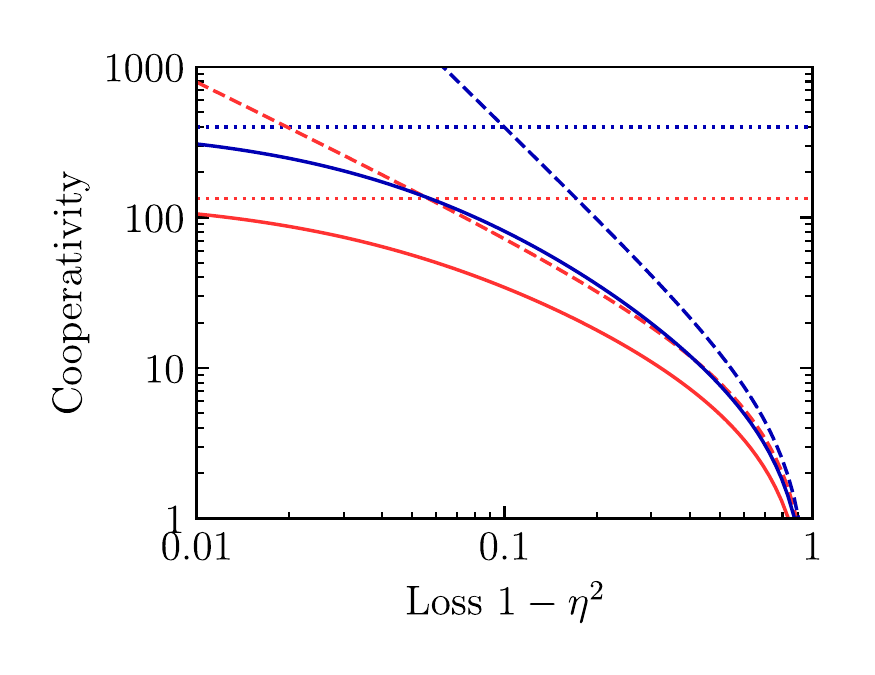}
\caption{Cooperativity as a function of optical power loss between systems. Cooperativity for the single-loop Eq.~\eqref{eq:cooperativity_1} (solid red line) and the double-loop Eq.~\eqref{eq:cooperativity_2} (solid blue line). Also shown are the maximum achievable cooperativities with zero losses (dotted lines). These amount to $4 c_1/(1/2+1/c_2)$ for the single loop and to $4 c_1 c_2$ for the double loop. For $c_2<1$, the cooperativities of these two geometries would almost coincide; for large $c_2$, they differ by a factor of $\sim c_2/2$. The dashed lines correspond to the limiting cases of infinite single-system cooperativities as given by Eqs.~\eqref{eq:cooperativity_1} for the single loop and \eqref{eq:cooperativity_2} for the double loop.}
\label{fig:cooperativity}
\end{figure}

In Fig.~\ref{fig:cooperativity}, we plot the resulting cooperativity of the light-mediated dynamics for the two schemes 1-2-1 and 1-2-1-2 as a function of the optical loss $1-\eta^2$. Here, we choose imbalanced systems with single-system cooperativities of $c_1 = 25$ and $c_2 = 4$. This imbalance is chosen in order to keep light-induced back-action on system 2 small compared to the coupling strength. Keeping $c_1 c_2$ constant and diminishing $c_2$ will asymptotically lead to the same cooperativity for the single loop and the double loop schemes. Remarkably, the figure shows that even for substantial optical losses of a few tens of percent, coherent light-mediated interactions between the two systems can be engineered.

\subsection{Sympathetic cooling}
\label{sec:cooling}

\begin{figure}[t]
\centering
\includegraphics[width=8.5cm]{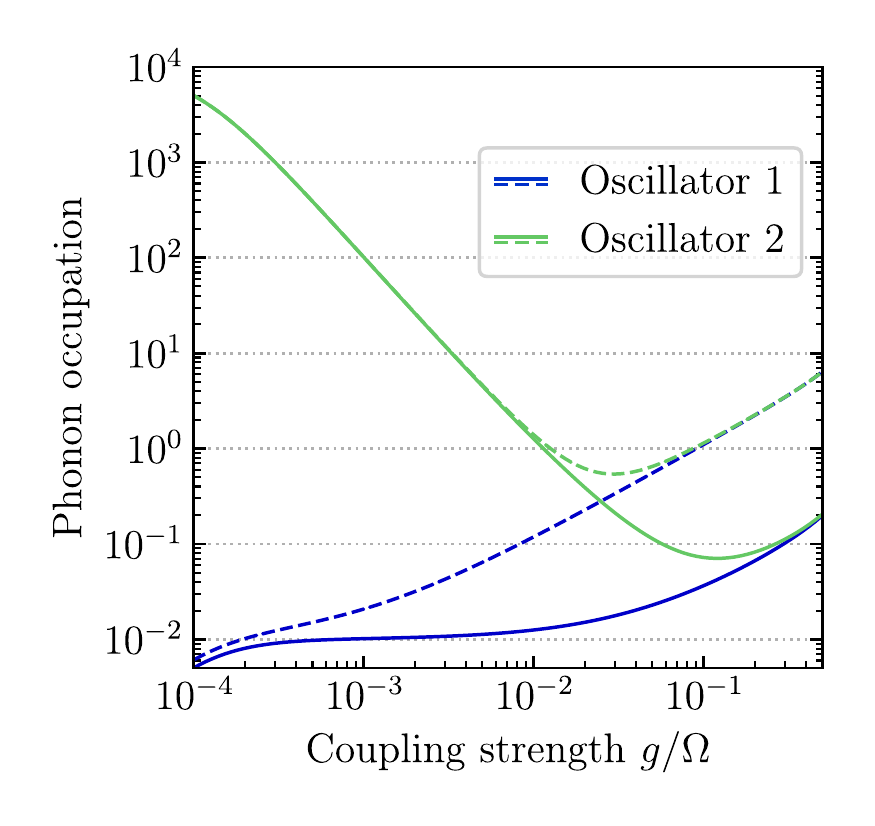}
\caption{Steady-state phonon numbers of both oscillators in the sympathetic cooling scenario described in the main text comparing the performance with losses ($\eta=0.9$, dashed lines) to the one without ($\eta=1$, solid lines). 
}
\label{fig:cooling}
\end{figure}

The probably simplest experiment that can be done using light-mediated Hamiltonian coupling is to study the thermalization of two oscillators in the presence of the beam-splitter coupling. Experimentally, this has been achieved in the context of sympathetic cooling of a mechanical oscillator coupled to the center-of-mass motion of a cloud of ultracold atoms \cite{Joeckel2015, Vochezer2018, Christoph2018}. These experimental setups are equivalent to Fig.~\ref{fig:schemes}(b) and we will focus on it in this section. In such a hybrid system the two oscillators exhibit fairly different characteristics. While the first oscillator (the ultracold atoms) is coupled to a vacuum bath ($\bar{n}_1\approx 0$) with large damping rate $\gamma_1$, the second oscillator (the mechanical oscillator) couples to a hot bath with a very low damping rate $\gamma_2$. Efficient sympathetic cooling of oscillator 2 via oscillator 1 occurs in a regime where the light-mediated coupling strength $g$ exceeds the thermal dissipation rate of oscillator 2 while remaining smaller than the damping rate $\gamma_1$ of oscillator 1. The minimum achievable phonon occupation is then limited by the cooperativity of Eq.~\eqref{eq:cooperativity_0} \cite{Vogell2013}. It is also evident from Eq.~\eqref{eq:lindblad_backaction_rwa} that back-action noise on the two oscillators increases their effective bath occupation, thus limiting the cooling efficiency. In the coupling geometry 1-2-1, the optimal strategy consists in choosing $c_1 \gg c_2 \approx 1$ such that the back-action rate $\Gamma_2$ on oscillator 2 remains insignificant compared to the cooling rate $\sim g^2/\gamma_1 = 2 \eta^2 c_1 \Gamma_2$ \cite{Vogell2013}.

In order to directly evaluate the steady-state phonon occupation of both oscillators, we treat the Gaussian dynamics of the coupled system using the covariance matrix formalism \cite{Hofer2015} (see Appendix \ref{sec:gaussian}). For the simulations, we choose two oscillators with equal frequency $\Omega$. Both interact with the light field via quantum-nondemolition (QND) interactions, i.e., $\theta_1 =\pi/4 = - \theta_2$ such that $\alpha = 1$ is maximal and no additional optical cooling of oscillator 2 occurs. Oscillator 1 has a large damping rate $\gamma_1 = 0.1\Omega$ that couples it to a vacuum bath ($\bar{n}_1=0$). Contrarily, oscillator 2 is connected to a hot bath with $\bar{n}_{2} = 10^4$ but its damping rate is very low ($\gamma_2 = 10^{-7}\Omega$) such that thermalization occurs at the comparably low rate $\gamma_{2,\mathrm{th}} = 10^{-3}\Omega \ll \gamma_1$. This is the typical situation encountered in hybrid atom-optomechanical systems \cite{Joeckel2015, Christoph2018}.

The resulting steady-state phonon occupations of the two oscillators in the sympathetic cooling scenario are plotted in Fig.~\ref{fig:cooling} as a function of the coherent coupling strength $g = 2\eta g_1 g_2$ keeping a fixed ratio of $g_1/g_2 = 10$. This ensures that $\Gamma_2 \ll g$. In the lossless case, cooling below unity phonon occupation of the mechanical oscillator is possible for a coupling strength of $g \approx \gamma_1 = 0.1 \Omega_{i}$. Increasing the coupling strength further leads to a breakdown of the simple cooling picture from above. As soon as $g >\gamma_1$, the modes hybridize which causes heating of oscillator 1 by oscillator 2. In the lossy case with $\eta=0.9$, substantial back-action heating of oscillator 1 leads to a much higher minimum phonon number in oscillator 2. Nevertheless, this value still lies below 1 indicating a certain robustness against losses.

\subsection{Entanglement}
\label{sec:entanglement}
As a second application, we consider the generation of entanglement between two oscillators, comparing the different looped and cascaded schemes of Fig.~\ref{fig:schemes}. Entanglement between two bosonic modes can be generated by the two-mode squeezing Hamiltonian $H_\mathrm{TMS}$ which produces squeezing in the quadratures $X_1 + X_2$ and $P_1 - P_2$ and anti-squeezing in $X_1 - X_2$ and $P_1 + P_2$, thus creating nonclassical correlations. In order to realize it using the looped cascaded interaction, we consider the mode $b_1$ to have negative frequency $\Omega_1 = -\Omega < 0$ and the mode $b_2$ to have positive frequency $\Omega_2 = \Omega > 0$. An inverted oscillator with $\Omega_1<0$ can directly be realized experimentally with a collective atomic spin pumped to its highest energy state \cite{Julsgaard2001} or, effectively, in cavity optomechanical systems driven by two optical tones \cite{Woolley2013, Ockeloen-Korppi2016}. In this setting $H_\mathrm{TMS}$, is stationary in the interaction picture and the steady-state two-mode squeezing parameter $r$ is approximately given by the ratio of all noise rates over the coherent coupling strength, i.e.,
\begin{equation}
r \approx \frac{\gamma_{1,\mathrm{th}} + \gamma_{2,\mathrm{th}} + (\Gamma_1 + \Gamma_2)/2}{2g\beta}.
\end{equation}
We see that the requirements for squeezing ($r<1$) are more restrictive than those for achieving large cooperativity because all decoherence rates need to be individually smaller than the coupling strength. In order to quantify the degree of entanglement, we evaluate two established non-separability criteria for Gaussian states, the logarithmic negativity \cite{Zyczkowski1998, Vidal2002} and the EPR variance \cite{Duan2000a,Simon2000} (see Appendix~\ref{sec:appendix_entanglement}).

\begin{figure*}[t]
\centering
\includegraphics[width=18cm]{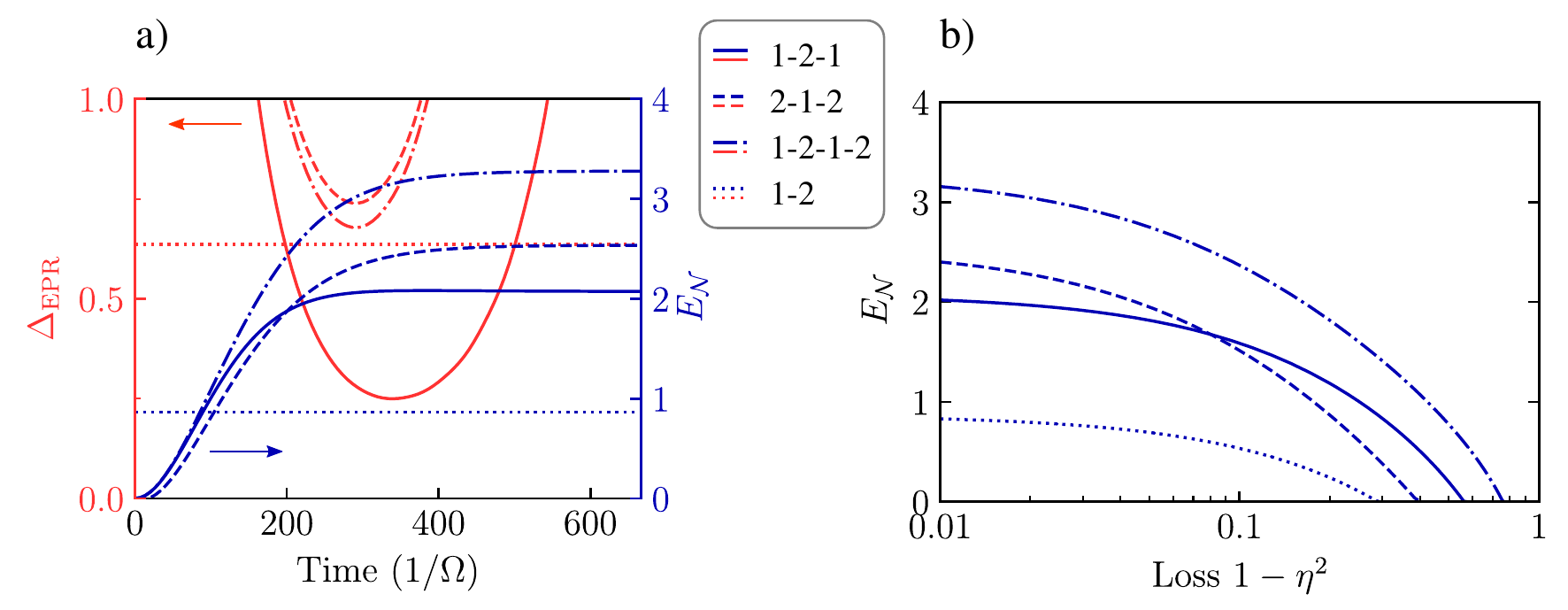}
\caption{Entanglement as characterized by the EPR variance (red color) and measured by the logarithmic negativity (blue color) in four relevant cases: (i) the looped geometry 1-2-1 (solid lines), (ii) the looped geometry with reverse order 2-1-2 (dashed lines), (iii) the double loop with interaction order 1-2-1-2 (dot-dashed lines), and (iv) the steady state of the simple cascaded scheme 1-2 (dotted lines). (a) Entanglement measures vs interaction time in the lossless case $\eta=1$. (b) Steady-state logarithmic negativity vs optical loss $1-\eta^2$.
}
\label{fig:entanglement}
\end{figure*}

In Fig.~\ref{fig:entanglement} we show the bipartite entanglement as quantified by EPR variance ($\Delta_\mathrm{EPR}<1$) and logarithmic negativity ($E_\mathcal{N}>0$) as a function of (a) the interaction time between the two oscillators and (b) the optical loss $1-\eta^2$. 
We plot them for four relevant cases: (i) the looped geometry of Fig.~\ref{fig:schemes}(b) with interaction order 1-2-1 (solid line), (ii) the looped geometry with reversed interaction order 2-1-2 (dashed line), and (iii) the double loop of Fig.~\ref{fig:schemes}(c) with interaction order 1-2-1-2 (dot-dashed line). For comparison, we also show (iv) the achievable steady-state entanglement using the simple cascaded scheme 1-2 of Fig.~\ref{fig:schemes}(a) (dotted lines). In the simulations, we deliberately choose a slight asymmetry in the damping rates and thermal bath occupations in order to describe the situation encountered in ongoing experiments in hybrid atom-optomechanics \cite{Joeckel2015, Moller2017}. While oscillator 1 couples to a vacuum bath with $\bar{n}_1 = 0$ and intrinsic decay rate $\gamma_1 = 10^{-3}\Omega$, oscillator 2 has a lower damping rate $\gamma_2 = 10^{-4}\Omega$ but a larger thermal occupation $\bar{n}_2 = 10$. For oscillator 1, we choose the QND light-matter interaction with $\theta_1 = \pi/4$ while for oscillator 2 a value of $\theta_2 = -0.8 \pi/4$ introduces an imbalance between beam-splitter and parametric gain interactions with the purpose of further cooling its thermal fluctuations. For the single-pass scheme, we set $\theta_2 = +0.8\pi/4$ such that both oscillators couple predominantly to the same optical quadrature \cite{Vasilyev2013, Huang2018}. The coupling constants $g_1$ and $g_2$ have been chosen in order to minimize the back-action rates while fixing $c_1 c_2 = 100$ and thus keeping $g$ constant. 
For the schemes 1-2 and 1-2-1-2, the choice is symmetric with $c_1 = c_2 = 10$. In either of the schemes 1-2-1 and 2-1-2, only one oscillator is protected from quantum noise and we increase the single-pass cooperativity of this one at the cost of the other. This leads us to the choices $c_1 = 25$, $c_2 = 4$ for 1-2-1 and $c_1 = 4$, $c_2 = 25$ for 2-1-2 such that the oscillator without back-action cancellation has a weaker coupling. 

In Fig.~\ref{fig:entanglement}(a) the dynamics start from an initial thermal state with $\bar{n}_1 = 0$ and $\bar{n}_2 = 10$. Strong entanglement is achieved after a short interaction time required to overcome the initial thermal noise in oscillator 2. The logarithmic negativities in the three cases (i)--(iii) reach steady states with similar values, with the double loop (iii) being optimal. This is a direct consequence of the efficient quantum noise cancellation on both systems, while in the single-loop schemes (i) and (ii) only a single system benefits from quantum noise cancellation. However, for cascaded systems with imbalanced single-pass cooperativities, these schemes are already close to optimal implying that the advantage of full back-action cancellation in (iii) can only be fully exploited if the experimentally achievable single-system cooperativities are high and if losses are low. 

Entanglement is demonstrated and quantified clearly in terms of the negativity. On top of that, we show the EPR variance, demonstrating how close the entangled state is to the ``canonical'' two-mode squeezed state. This is a relevant question regarding applications of the entangled state for quantum information protocols such as teleportation \cite{Braunstein2005}.
We see that the performance of the different schemes in terms of EPR entanglement is quite different from that in terms of the negativity. While scheme 1-2-1 attains a minimum of $\Delta_\mathrm{EPR} \approx 0.25$, the schemes 2-1-2 and 1-2-1-2 only achieve weak squeezing of the EPR variance. This behavior can be understood from the strong imbalance of the thermal and back-action noise processes acting on the two oscillators that leads to a deviation of the squeezed quadratures from $X_1 + X_2$ and $P_1 - P_2$. In the schemes 1-2 and 1-2-1, however, optical back-action cooling of oscillator 2 reduces its thermal noise and increases its damping rate, thereby partially lifting the imbalance and reducing $\Delta_\mathrm{EPR}$. This mechanism is absent in the other schemes 2-1-2 and 1-2-1-2 where optical back-action on oscillator 2 cancels. Moreover, all schemes merely show transient EPR entanglement as for long interaction times growing noise in the anti-squeezed quadrature enters $X_1 + X_2$ and $P_1 - P_2$ and leads to an increase of $\Delta_\mathrm{EPR}$. However, we emphasize that all schemes do indeed achieve entanglement in the stationary state as witnessed by the logarithmic negativity.

As compared to steady-state entanglement that would be achieved in the simple cascaded scheme 1-2 with identical light-matter interactions, the looped geometries perform better. We note, however, that entanglement in the simple cascaded scenario can in principle be further optimized by additional tuning of the local light-matter interactions \cite{Huang2018}. Another advantage of the coherent entanglement achieved via the looped geometries is a faster entanglement rate that does not rely on reaching a steady state after a much longer interaction time.

Finally, in Fig.~\ref{fig:entanglement}(b), we analyze the dependence of the achievable entanglement on optical losses. Here we only show logarithmic negativity because the entanglement of bipartite Gaussian states is solely determined by two-mode squeezing \cite{Duan2000a, Kraus2003}. Remarkably, all schemes are very robust against losses of up to 10\% without strong degradation and still show $E_\mathcal{N}>0$ until $1-\eta^2 = 40\%$ for 1-2 and 2-1-2, 50\% for 1-2-1, and even 70\% for 1-2-1-2.

\subsection{Unconditional squeezing of a single oscillator}

Having discussed the coherent dynamics between two distinct quantum systems interacting via an optical mode, we devote this last section to engineering coherent dynamics in a single quantum system. If we consider the scheme in Fig.~\ref{fig:schemes}(b) and leave out system 2, the remaining dynamics of system 1 alone are very interesting on its own. Multipass interactions between light and atomic ensembles have been subject to several theoretical studies investigating quantum memory and atom-light entanglement \cite{Muschik2006} or spin squeezing \cite{Trail2010, Wang2017}.

Here, we investigate the effect of a phase shift $\phi$ on the light field quadratures in between the two light-matter interactions. The discussion in Sec.~\ref{sec:looped_interaciton} already revealed the two effects: There is a light-mediated self-interaction and interference of back-action noise. For $\phi=\pi$, one has full back-action cancellation without self-interaction, meaning that even though the system strongly interacts with light, there is no effect visible to an external observer. For intermediate optical phase shift $\phi \in (0,\pi)$, the system is driven by itself, thus representing a case of coherent quantum feedback \cite{Guimond2017, Zhang2017}. The corresponding master equation is
\begin{eqnarray*}
\dot{\rho} &=& \ui \eta g_1^2 \sin \phi [B_1^\dag B_1, \rho] + 2 g_1^2 (1 + \eta \cos \phi) \mathcal{D}[B_1] \rho.
\end{eqnarray*}
For $B_1 = X_1$ being a harmonic oscillator quadrature, the effective Hamiltonian is equivalent to a one-axis twisting Hamiltonian \cite{Kitagawa1993} implementing squeezing at rate $g = \eta g_1^2 \sin\phi$. However, there is simultaneous back-action noise at rate $\Gamma_1 = 2 g_1^2 (1 + \eta \cos\phi)$ which does not cancel for any non-vanishing $g$. Nevertheless, the ratio of noise rate over squeezing rate
\begin{equation}
r = \frac{\Gamma_1/2}{g} = \frac{1 + \eta \cos\phi}{\eta \sin\phi},
\end{equation}
can be minimized for a given loss, leading to a value of $\phi$ close to $\pi$ where $r\sim (1-1/\eta)/(\phi-\pi)$. This scheme can in principle achieve arbitrarily large squeezing provided that $g$ remains large compared to other intrinsic and technical decoherence rates. The ratio $r$ does not give a bound for the minimum achievable squeezing parameter but rather expresses how much excess noise is added to the anti-squeezed quadrature. Besides applications in spin-squeezing similar schemes could equally well be employed to achieve squeezing of a mechanical oscillator in an optomechanical cavity. To this end, one has to recycle the cavity output field by first applying the relative phase shift $\phi$ between the local oscillator and the quantum field, and then sending it back into the cavity on a different mode.

Recently, \citet{Wang2017} have shown that one can erase the remaining back-action of the two-pass scheme in a three-pass configuration to achieve unitary spin-squeezing of an atomic ensemble. With phase shifts $\phi_{ij}$ between passes $i$ and $j$, the full master equation in this case reads
\begin{eqnarray*}
\dot{\rho} &=& - \alpha \;g_1^2 \; [X_1, X_1\rho] + \mathrm{h.c.},
\end{eqnarray*}
where $\alpha = \frac{3}{2} + \eta e^{-\ui\phi_{12}} + \eta e^{-\ui\phi_{23}} + \eta^2 e^{-\ui(\phi_{12}+\phi_{23})}$. For full noise cancellation at $\eta = 1$ one needs to solve 
\begin{equation*}
\re{\alpha}=\frac{3}{2} + \cos(\phi_{12}) + \cos(\phi_{23}) + \cos(\phi_{12}+\phi_{23}) \stackrel{!}{=} 0.
\end{equation*}
Setting $\phi_{12} = \phi_{23} = \phi$, the solution is found to be $\phi = \pm 2\pi/3$. The coherent interaction strength is given by $\im\alpha = \mp \sqrt{3}/2$. Using this choice of $\phi$ for $\eta<1$, we get
\begin{equation}
\dot{\rho} = - \ui \frac{\sqrt{3}(2\eta-\eta^2)}{2}g_1^2 [X_1^2, \rho] + (3-2\eta-\eta^2)g_1^2  \mathcal{D}[X_1]\rho.
\end{equation}
Here, one obtains a more favorable ratio of
\begin{equation}
r = \frac{\Gamma_1/2}{g} \sim \frac{4(1-\eta)}{\sqrt{3}\eta},
\end{equation}
which can be smaller than $0.1$ up to a power loss per pass of $1-\eta^2 \approx 8\%$.

\section{Conclusion}

Cascaded quantum systems have so far only been considered for dissipative entanglement schemes or unidirectional quantum communication. Here, we have extended this framework to include multiple interactions of an optical field with the individual quantum systems. In this case, light can also mediate coherent interactions between the quantum systems without adding noise to them. For two cascaded quantum systems, this is achieved if the light field interacts twice with the systems and if the second interaction with each of them is the time reversal of the first. In this situation, coherent driving of each system by the other is accompanied by a destructive interference of measurement back-action noise due to the light field, thus realizing an ideal Hamiltonian coupling. 

In order to quantify the strength of the coherent light-mediated interaction in the presence of experimental imperfections we have defined a cooperativity as the ratio of the coherent coupling constant over intrinsic and light-induced dissipation rates. Importantly, we have shown that large cooperativity can be achieved even in the presence of significant optical loss that renders the back-action cancellation imperfect. This robustness is very appealing for experiments and we believe that future quantum networks will benefit from the possibilities opened up by Hamiltonian interactions across macroscopic distances. Our scheme is particularly suited to interface hybrid quantum systems with distinct physical properties for which we have demonstrated its potential for ground-state sympathetic cooling and strong two-mode squeezing.

Since the looped cascaded interaction necessarily erases all information about the interacting systems on the optical field, one needs to find an alternative measurement strategy in order to verify the coupled dynamics at the quantum level. One simple solution could be auxiliary readout modes for each system, which would however make experimental realizations more complicated. As a more direct approach, we imagine real-time control of the optical field in order to switch from coherent dynamics to a collective measurement. Simultaneous weak measurement and partial noise cancellation are directly implemented in the cascaded scheme 1-2-1 with a loop on a single system. This presents an interesting intermediate scenario where one could explore the interplay of coherent dynamics and conditional quantum state evolution.


\begin{acknowledgements}
This work was supported by the project ``Modular mechanical-atomic quantum systems'' (MODULAR) of the European Research Council (ERC) and by the Swiss Nanoscience Institute (SNI).
\end{acknowledgements}

\appendix

\section{Proof that $\Lambda_\mathrm{eff}$ is positive semidefinite}
\label{sec:proof_lambdaeff_positive}
Here we prove that $\Lambda_\mathrm{eff}$ as defined in Eq.~\eqref{eq:def_lambdaeff} is always positive semidefinite. Applying this definition to Eq.~\eqref{eq:a_operator_loss} we rewrite
\begin{eqnarray*}
\Lambda_\mathrm{eff} &=& \sum_{i=1}^{n}\sum_{j<i} \eta_{ij} g_i g_j  (B_i^\dag B_j + B_j^\dag B_i) + \sum_{i=1}^{n} g_i^2  B_i^\dag B_i\\
&=& \mathbf{B}^\dag (\mathbf{g}\mathbf{g}^{T} \circ M_n) \mathbf{B},
\end{eqnarray*}
where we defined $\mathbf{B} = (B_1, \ldots, B_n)^T$, $\mathbf{g} = (g_1, \ldots, g_n)^T$ and 
\begin{equation*}
(M_n)_{ij} = \begin{cases}
\eta_{ij} & \text{for } i\neq j\\
1 & \text{for } i=j.
\end{cases}
\end{equation*}
The symbol $\circ$ denotes the Hadamard product of two matrices. If the product $\mathbf{g}\mathbf{g}^{T} \circ M_n\geq0$ then also $\Lambda_\mathrm{eff}\geq 0$. Since $\mathbf{g}\mathbf{g}^{T}$ is positive it remains to show that $M_n$ is positive semidefinite for all $n\geq1$ (Schur product theorem). We can construct recursively
\begin{equation*}
M_{n+1} = \left(\begin{array}{c|c}
M_n & \mathbf{a}_n\\\hline
\mathbf{a}_n^T & 1
\end{array}\right),
\end{equation*}
using the vector $\mathbf{a}_n = \eta_n (\mathbf{a}_{n-1}^T,1)^T$ with $\mathbf{a}_1 = \eta_1$ and $M_1=1$. We need to show that for any vector $\mathbf{v}\in \mathbb{R}^{n+1}$ and any $n\geq 0$ the expression $\mathbf{v}^T M_{n+1}\mathbf{v} \geq 0$. Decomposing $\mathbf{v} = (\mathbf{w}^T, x)^T$ where $\mathbf{w}\in \mathbb{R}^{n}$ and $x\in\mathbb{R}$ we have
\begin{eqnarray*}
\mathbf{v}^T M_{n+1}\mathbf{v} &=&  \mathbf{w}^T M_{n}\mathbf{w} + 2 x \mathbf{w}^T \mathbf{a}_n + x^2\\
& = & \mathbf{w}^T M_{n}\mathbf{w} - (\mathbf{w}^T \mathbf{a}_n)^2 + (x+\mathbf{w}^T \mathbf{a}_n)^2\\
&\geq& \mathbf{w}^T M_{n}\mathbf{w} - (\mathbf{w}^T \mathbf{a}_n)^2\\
&=& \mathbf{w}^T (M_{n} - \mathbf{a}_n \mathbf{a}_n^T)\mathbf{w}.
\end{eqnarray*}
It follows that $M_{n+1} \geq 0$ if $M_n\geq\mathbf{a}_n \mathbf{a}_n^T$. To show the latter we note that $\mathbf{a}_n \mathbf{a}_n^T = \eta_n^2 \mathbf{b}_n \mathbf{b}_n^T \leq \mathbf{b}_n \mathbf{b}_n^T$ with $\mathbf{b}_n = (\mathbf{a}_{n-1}^T,1)^T$. Since 
\begin{equation*}
M_{n} - \mathbf{b}_n \mathbf{b}_n^T = \left(\begin{array}{c|c}
M_{n-1} - \mathbf{a}_{n-1} \mathbf{a}_{n-1}^T& \mathbf{0}\\\hline
\mathbf{0} & 0
\end{array}\right),
\end{equation*}
it follows that $M_{n} \geq \mathbf{a}_n \mathbf{a}_n^T$ if $M_{n-1} \geq \mathbf{a}_{n-1} \mathbf{a}_{n-1}^T$. Since $M_1 = 1 \geq \eta_1^2 = \mathbf{a}_{1} \mathbf{a}_{1}^T$ the proof follows by induction.

\section{Master equation}
\label{sec:appendix_master_equation}

This section aims to show how the master equation can be transformed into Lindblad form. Starting from the general master equation~\eqref{eq:master_eq_general_form} we expand $A = \sum_{i,j} A_{ij} B_i^\dag B_j$ and $\mathcal{J}\rho = \sum_{i,j} A_{ij} B_j \rho B_i^\dag + A_{ij}^\ast B_i \rho B_j^\dag$. Then,
\begin{equation}
\dot{\rho} = - \sum_{i,j} A_{ij} [B_i^\dag, B_j \rho] + \mathrm{h.c.} 
\label{eq:master_eq_general}
\end{equation}
We now identify Hamiltonian and dissipative part of $A$ by the relations $R_{ij} = -\ui \hbar (A_{ij} - A_{ji}^\ast)$ and $L_{ij} = A_{ij} + A_{ji}^\ast$ equivalent to Eqs.~\eqref{eq:general_eff_hamiltonian} and \eqref{eq:def_lambdaeff}, respectively. In this basis the effective Hamiltonian and effective dissipation read $H_\mathrm{eff}= \frac{1}{2}\sum_{i,j} R_{ij} B_i^\dag B_j$ and $\Lambda_\mathrm{eff}= \sum_{i,j} L_{ij} B_i^\dag B_j$. Using this notation it follows that
\begin{eqnarray}
\dot{\rho} &=& - \frac{\ui}{\hbar} \sum_{i,j} \frac{R_{ij}}{2} [B_i^\dag B_j, \rho] \\
& &  - \sum_{i,j} \frac{L_{ij}}{2} (B_i^\dag B_j \rho + \rho B_i^\dag B_j - 2 B_j \rho B_i^\dag),
\end{eqnarray}
because both $R_{ij}^\ast = R_{ji}$ and $L_{ij}^\ast = L_{ji}$. Provided that $L$ is positive semidefinite the master equation is physical. By diagonalizing $L = \sum_{k} \gamma_k \mathbf{e}_k \mathbf{e}_k^\dag$ with eigenvalues $\gamma_k \geq 0$ and orthonormal eigenvectors $\mathbf{e}_k$ we define the (unnormalized) jump operators $j_k = \sqrt{\gamma_k}\mathbf{e}_k^\dag \mathbf{B}$. The eigenvalues are the corresponding dissipation rates. Using this procedure we obtain the Lindblad form \eqref{eq:master_eq_general_lindblad}.

\section{Time delays}
\label{sec:time_delays}

\subsection{Master equation}
Here, we model the effect of time delays on the master equation. We work in an interaction picture with respect to $H_0$ including local decoherence. For the ladder operators we have $\tilde{b}_j(t) = b_j e^{-\ui\Omega_jt - \gamma_jt/2}$ where $b_j$ is the corresponding operator in the Schr\"odinger picture. Thus,
\begin{eqnarray*}
\tilde{B}_j(t - \tau) &=& (\mu_j \tilde{b}_j(t) e^{\ui\Omega_j \tau} + \nu_j \tilde{b}^\dag_j(t) e^{-\ui\Omega_j \tau})e^{\gamma_j\tau/2}\\
&=& (\tilde{B}_j(t) \cos(\Omega_j \tau) + \ui \tilde{B}^{-}_j(t) \sin(\Omega_j \tau))e^{\gamma_j\tau/2},
\end{eqnarray*}
where we defined $\tilde{B}^{-}_j = \mu_j \tilde{b}_j - \nu_j \tilde{b}_j^\dag$. Typically, the decay rates $\gamma_j$ are considered to be much smaller than the oscillation frequencies $\Omega_j$. In this case, the approximation to neglect delays is based on the smallness of the parameter $\epsilon_j = \Omega_j\tau$. To first order and setting $e^{\gamma_j\tau/2}\approx1$ we thus have $\tilde{B}_j(t - \tau) \approx \tilde{B}_j + \ui\epsilon_j \tilde{B}_j^{-}$. Including these delays we find that the operator $A$ of the general master equation becomes
\begin{eqnarray}
A	&=& \sum_{k<j} \eta_{jk} g_j g_k \tilde{B}_j^\dag(t) \tilde{B}_k(t - \tau_{jk}) + \sum_{j} \frac{g_j^2}{2}  \tilde{B}_j^\dag(t) \tilde{B}_j(t)\nonumber\\
	&\approx& \sum_{k<j} \eta_{jk} g_j g_k  \tilde{B}_j^\dag \tilde{B}_k + \sum_{j} \frac{g_j^2}{2}  \tilde{B}_j^\dag \tilde{B}_j\\
	& & + \;\ui \sum_{k<j}  \epsilon_{jk} \eta_{jk} g_j g_k \tilde{B}_j^\dag \tilde{B}_k^{-}, \label{eq:A_delay_correction}
\end{eqnarray}
where we have defined $\tau_{jk} = \tau_j - \tau_k$ and $\epsilon_{jk} = \Omega_k\tau_{jk}$. The line \eqref{eq:A_delay_correction} thus presents a correction to the master equation due to delays. In order to neglect it altogether we have to compare the associated rate $\eta_{jk} g_j g_k \epsilon_{jk}$ to all other rates in the system, in particular the smallest rates are the damping rates $\gamma_j$. Defining the quality factor $Q_j = \Omega_j/\gamma_j$ we obtain the condition $\eta_{jk} g_j g_k \ll Q_j^{-1}\tau_{jk}^{-1}$ which can be a fairly restrictive upper bound for the light-mediated coupling strengths if the damping rates are small and the delays are large. Instead of this heuristic argument a more rigorous stability criterion has to be applied in general \cite{Vochezer2018} which however goes beyond the scope of this article.


\subsection{Back-action cancellation}

We derive here the equation of motion for the first oscillator in the looped scenario 1-2-1 in the presence of time delays
\begin{eqnarray*}
\dot{b}_1 &=& (-\ui\Omega - \gamma_1'/2)b_1 + g_1 \left(\mu_1 f_\mathrm{in,1} + \nu_1 f_\mathrm{in,1}^\dag\right)\\
 & & - \eta_2 g_1 g_2 e^{-\ui\phi} \left(\mu_1 B_2(t-\tau_{12}) - \nu_1 B_2^\dag(t-\tau_{12})\right),
%
\end{eqnarray*}
with optically modified damping rate $\gamma_1'$ (cf. Appendix~\ref{sec:gaussian}) and optical back-action force
\begin{eqnarray*}
f_\mathrm{in,1} &=& a_\mathrm{in}(\zeta_1) + e^{-\ui\phi} \Big [\eta_2 \eta_1 a_\mathrm{in}(\zeta_3) + \sqrt{1-\eta_1^2\eta_2^2} h_\mathrm{in}(\zeta_3)\Big],
\end{eqnarray*}
with combined noise input $h_{\mathrm{in}}$ due to losses. The back-action noise driving $b_1$ is filtered with its susceptibility $\chi_1(\omega) = [\gamma_1'/2 - \ui(\omega- \Omega_1)]^{-1}$. Within the small bandwidth $\gamma_1\ll\Omega_1$ one can then approximate
\begin{equation*}
f_\mathrm{in,1} \approx \int \frac{\ud\omega}{\sqrt{2\pi}}a_\mathrm{in}(\omega)e^{\ui\omega\tau_1}(1 + \eta_1\eta_2 e^{-\ui\phi} e^{\ui\Omega_1\tau_{13}})  + \ldots,
\end{equation*}
omitting the noise term involving the uncorrelated input $h_{\mathrm{in}}$. This leads to a suppression factor for $\phi=\pi$ of $1 - \eta_1\eta_2 e^{\ui\Omega_1\tau_{13}}$ for both $a_\mathrm{in}$ and $a_\mathrm{in}^\dag$. Significant delay introduces a phase shift that adds back-action from the orthogonal optical quadrature. Consequently, the total back-action force amounts to
\begin{eqnarray*}
\dot{b}_1 &\approx& g_1 (1-\eta_1\eta_2 \cos(\Omega_1\tau_{13}))\left(\mu_1 a_\mathrm{in} + \nu_1 a_\mathrm{in}^\dag\right)\\
 & & -\ui g_1\eta_1\eta_2 \sin(\Omega_1\tau_{13}))\left(\mu_1 a_\mathrm{in} - \nu_1 a_\mathrm{in}^\dag\right)\\
 & & - g_1 \sqrt{1-\eta_1^2\eta_2^2} \left(\mu_1 h_\mathrm{in} + \nu_1 h_\mathrm{in}^\dag\right).
\end{eqnarray*}
We also see that changing the phase shift to $\phi = \pi + \Omega_1\tau_{13}$ allows us to compensate for the delay and recover full back-action cancellation.

\section{Gaussian Dynamics}
\label{sec:gaussian}

Starting from the master equation \eqref{eq:master_eq_general} we can derive equations of motion for expectation values of any system operator. Assuming the system operators are linear, we first transform $B_i = \sum_{j} U_{ij} Q_j$ into a basis of canonical operators $Q_i$ with $U$ being the basis transformation matrix. The operators $Q_i$ are Hermitian and satisfy the commutation relation $[Q_i, Q_j] = \ui J_{ij}$ with $J$ being a real skew-symmetric matrix as for standard harmonic oscillators. The matrix $A$ transforms under $U$ into $\tilde{A} = U^\dag A U$. We then obtain the transformed master equation
\begin{equation*}
\dot{\rho} = - \sum_{i,j} \tilde{A}_{ij} [Q_i, Q_j \rho] + \mathrm{h.c.} 
\end{equation*}

The time evolution of the expectation value of any system operator $\bar{Y} = \langle Y \rangle$ reads\begin{IEEEeqnarray}{r C l}
\frac{\ud}{\ud t} \bar{Y} &=& \Tr\{Y \dot{\rho}\}\\
&=& - \sum_{i,j} \left(\tilde{A}_{ij} \langle [Y, Q_i] Q_j\rangle - \tilde{A}^\ast_{ij} \langle Q_j [Y,Q_i] \rangle\right).\nonumber
\end{IEEEeqnarray}

For first and second moments closed-form equations of motion can be derived. We define the symmetric covariance matrix as
\begin{equation}
\bar{C}_{kl} = \frac{1}{2} \langle Q_k Q_l + Q_l Q_k\rangle - \langle Q_k\rangle \langle Q_l\rangle.
\label{eq:symm_covariance}
\end{equation}
The equations of motion read
\cite{Edwards2005}
\begin{IEEEeqnarray}{r C l}
\frac{\ud}{\ud t} \bar{\mathbf{Q}} &=& F \bar{\mathbf{Q}},\label{eq:eom_1stmoment}\\
\frac{\ud}{\ud t} \bar{C} &=& F \bar{C} + \bar{C} F^T + N. \label{eq:eom_2ndmoment}
\end{IEEEeqnarray}
Here we defined the real-valued matrices $F$ and $N$ as
\begin{eqnarray}
F &=& 2 J \im\{\tilde{A}\},\\
N &=& J \re\{\tilde{A} + \tilde{A}^T\} J^T,
\end{eqnarray}
which describe drift and diffusion, respectively, of the Gaussian state. In terms of the Hamiltonian and dissipative parts of $A$, $R$ and $L$, respectively, these can be re-written as
\begin{eqnarray}
F &=& J\left(\re\{\tilde{R}\} + \im\{\tilde{L}\}\right),\\
N &=& J\re\{\tilde{L}\}J^T.
\end{eqnarray}
Equation~\eqref{eq:eom_2ndmoment} is used to calculate the entanglement dynamics in the looped schemes in Sec.~\ref{sec:entanglement}.

In steady state $\bar{\mathbf{Q}} = 0$ and the covariances are obtained by solving the Lyapunov equation
\begin{equation}
F \bar{C} + \bar{C} F^T + N = 0.
\end{equation}
This equation is solved in order to obtain the steady-state phonon occupations in Sec.~\ref{sec:cooling} and the steady-state entanglement in the simple cascaded scenario 1-2 in Sec.~\ref{sec:entanglement}.

In the following we assume two harmonic oscillators such that $Q = (X_1, P_1, X_2, P_2)$ with $[X_i, P_j] = \ui \delta_{ij}$. For the purpose of illustration we first consider the looped geometry 1-2-1 for $\phi = \pi$, $\eta_1 = \eta_2 = \eta$ and local QND interactions with $B_1 = \ui X_1$ and $B_2 = X_2$. This corresponds to a coupling Hamiltonian $H_\mathrm{eff} \propto X_1 X_2$. The drift matrix then evaluates as
\begin{equation}
F = \begin{pmatrix}
-\gamma_1/2			& \Omega_1		& 0 			& 0\\
-\Omega_1			& -\gamma_1/2	& - g 			& 0\\
0					& 0				& -\gamma_2/2 	& \Omega_2\\
- g					& 0				& -\Omega_2 	& -\gamma_2/2
\end{pmatrix},
\end{equation}
and the diffusion matrix is
\begin{equation}
N = \begin{pmatrix}
\gamma_{1,\mathrm{th}}	& 0	& 0 & 0\\
0	& \gamma_{1,\mathrm{th}} + \Gamma_1 & 0 & 0\\
0	& 0	& \gamma_{2,\mathrm{th}}	& 0\\
0 	& 0	& 0	& \gamma_{2,\mathrm{th}} + \Gamma_2
\end{pmatrix},
\end{equation}
with $g=2\eta g_1 g_2$, $\Gamma_1 = 2 g_1^2 (1 - \eta^2)$ and $\Gamma_2 = g_2^2$.

In order to solve the equation of motion \eqref{eq:eom_2ndmoment} of the covariance matrix we assume an initial thermal state $\bar{C}_0 = \mathrm{diag}\left(\bar{n}_{\mathrm{th},1} + \frac{1}{2},\bar{n}_{\mathrm{th},1} + \frac{1}{2},\bar{n}_{\mathrm{th},2} + \frac{1}{2},\bar{n}_{\mathrm{th},2} + \frac{1}{2}\right)$.

In Sec.~\ref{sec:entanglement} we consider more general light-matter interactions $B_i = \mu_i b_i + \nu_i b_i^\dag$ with $\mu_i = \cos\theta_i$ and $\nu_i = \sin\theta_i$. We find modified damping rates $\gamma_i' = \gamma_i + \cos(2\theta_i)\Gamma_i$ and back-action rates $\Gamma_{i,X} = \frac{1-\sin(2\theta_i)}{2} \Gamma_i$ and $\Gamma_{i,P} = \frac{1+\sin(2\theta_i)}{2} \Gamma_i$. With these changes the drift matrix for the 1-2-1 scheme then reads
\begin{widetext}
\begin{equation}
F = \begin{pmatrix}
-\gamma_1'/2		& \Omega_1		& g (\alpha - \beta)/2 			& 0\\
-\Omega_1			& -\gamma_1'/2	& 0 			& g (\alpha + \beta)/2\\
- g	(\alpha + \beta)/2					& 0	& -\gamma_2'/2 	& \Omega_2\\
0 & -g (\alpha - \beta)/2				& -\Omega_2 	& -\gamma_2'/2
\end{pmatrix},
\end{equation}
and the diffusion matrix is
\begin{equation}
N = \begin{pmatrix}
\gamma_{1,\mathrm{th}} + \Gamma_{1,X}	& 0	& 0 & 0\\
0	& \gamma_{1,\mathrm{th}} + \Gamma_{1,P} & 0 & 0\\
0	& 0	& \gamma_{2,\mathrm{th}} + \Gamma_{2,X}	& 0\\
0 	& 0	& 0	& \gamma_{2,\mathrm{th}} + \Gamma_{2,P}
\end{pmatrix}.
\end{equation}
\end{widetext}
In the double-loop scenario 1-2-1-2, back-action rates change to $\Gamma_2= 2 g_2^2 (1-\eta^2)$ and coupling strengths in the lower triangle are multiplied by $2-\eta^2$ for the additional pass through system 2. Moreover, the diffusion matrix acquires small off-diagonal entries for the covariances of $X_1, X_2$ and $P_1, P_2$ because of an increased uni-directionality for finite loss.


\section{Gaussian state entanglement criteria}
\label{sec:appendix_entanglement}


In order to quantify the degree of entanglement between two oscillators in Sec.~\ref{sec:entanglement} we evaluate two established non-separability criteria for Gaussian states. The first one is the logarithmic negativity \cite{Zyczkowski1998, Vidal2002}
\begin{equation}
E_\mathcal{N} = \sum_{\pm} \max\left(0, - \log_2\left(2\tilde{c}_\pm\right)\right),
\end{equation}
where $\tilde{c}_\pm = \sqrt{(p \pm \sqrt{p^2 - 4 q})/2}$ are the symplectic eigenvalues of the partial transpose of the covariance matrix $\bar{C}^{T_1}$. Defining the block-matrix form of the $4\times4$ covariance matrix,
\begin{equation}
\bar{C} = \begin{pmatrix}
v_1 &v_{12}\\
v_{12}^T & v_2
\end{pmatrix},
\end{equation}
the coefficients evaluate as $p = \det{v_1} + \det{v_2} - 2\det{v_{12}}$ and $q=\det{\bar{C}}$. The logarithmic negativity directly measures the two-mode squeezing parameter $r$ as $E_\mathcal{N} \approx - \log_2(r)$ \cite{Kraus2003}. 

As a second entanglement criterion we evaluate the EPR variance \cite{Duan2000a,Simon2000}
\begin{equation}
\Delta_\mathrm{EPR} = \frac{1}{2}\left[\Var(X_1 + X_2) + \Var(P_1 - P_2)\right],
\end{equation}
which is conveniently expressed in terms of experimentally accessible variances. The EPR variance detects entanglement for $\Delta_\mathrm{EPR}<1$ and even stronger EPR correlations for $\Delta_\mathrm{EPR}<0.5$ \cite{Reid2009}.
We remark that further local unitary transformations of the quadratures $X_i$, $P_i$ and a relative weighting between systems 1 and 2 would have to be included in $\Delta_\mathrm{EPR}$ in order to obtain not only a sufficient, but also a necessary criterion for entanglement \cite{Duan2000a}.


\bibliography{light_mediated_interactions_bib}

\end{document}